\documentclass[draftclsnofoot,onecolumn, 12pt]{IEEEtran}
\usepackage{color}
\usepackage{graphicx, pstool}
\usepackage{amsmath}
\usepackage{amsfonts}
\usepackage{amssymb}
\usepackage{booktabs}
\usepackage{dsfont}
\usepackage{cases}
\usepackage{tabularx}
\usepackage{boxedminipage}
\usepackage{url}
\usepackage{mathrsfs}

\begin{document}

\title{  On the Capacity of the Two-Hop Half-Duplex Relay Channel}
\author{\normalsize Nikola Zlatanov, Vahid Jamali, Derrick Wing Kwan Ng, and   Robert
Schober
\thanks{This work has been accepted for presentation in part at  IEEE Globecom 2015}
\thanks{N. Zlatanov is  with the Department of Electrical and Computer Systems Engineering, Monash University, Melbourne, VIC 3800, Australia (e-mail: nikola.zlatanov@monash.edu).}
\thanks{ V. Jamali and R. Schober are with the Friedrich-Alexander University of Erlangen-N\"urnberg,
Institute for Digital Communications, D-91058 Erlangen, Germany
(e-mails: vahid.jamali@fau.de  robert.schober@fau.de).
}
\thanks{ D. W. K. Ng  is with the School of Electrical Engineering and Telecommunications, University of New South Wales, Sydney,
N.S.W. 2052, Australia (e-mail: w.k.ng@unsw.edu.au).}
}

\maketitle

\begin{abstract}
Although extensively investigated, the capacity of the two-hop half-duplex (HD) relay channel is not fully    understood. In particular, a capacity  expression which can be easily evaluated  is not available and an explicit coding scheme which achieves the capacity is not known either. In this paper,   we derive a new expression for the   capacity of the two-hop HD relay channel by simplifying previously derived converse expressions. Compared to previous results,  the new capacity expression can be easily evaluated. Moreover, we propose an explicit coding scheme which  achieves the capacity. To achieve the capacity,    the relay does not only send  information   to the destination by   transmitting  information-carrying symbols but also      with the zero symbols resulting from the relay's silence during reception.
As   examples, we compute the capacities of the two-hop HD relay channel  for the cases when  the source-relay and  relay-destination links are  both binary-symmetric channels (BSCs) and additive white Gaussian noise (AWGN) channels, respectively, and   numerically  compare  the capacities  with the rates achieved by conventional relaying where the relay receives and transmits  in a codeword-by-codeword fashion and switches between  reception and transmission in a strictly alternating manner. Our numerical results  show that the capacities of the two-hop HD relay channel  for BSC and AWGN links are significantly larger than the rates achieved with conventional relaying.
\end{abstract}


\IEEEpeerreviewmaketitle

\newtheorem{theorem}{Theorem}
\newtheorem{corollary}{Corollary}
\newtheorem{remark}{Remark}
\newtheorem{example}{Example}
\newtheorem{lemma}{Lemma}

\section{Introduction}

The relay channel is one of the building blocks of any general network. As such, it has been widely investigated in the literature e.g. \nocite{cover,  1435648, cheap_paper,kramer2004models, 6763001, kramer_book_1, meulen, laneman1,laneman2, 5730555, 1321222, 1614082, 1397921, 1214069, 4294174, 4373409, 4107956, 1542409, 1499041, 6121995} \cite{cover}-\cite{6121995}. The simplest relay channel is the
  two-hop relay channel   comprised of a source, a relay, and a destination, where the direct link between the source and the destination is not available. In this relay channel,  the source transmits a message  to the relay, which then forwards it to the destination. Generally, a relay  can employ two different modes of reception and transmission, i.e.,   full-duplex (FD)   and   half-duplex (HD). In the FD mode, the relay   receives and transmits at the same time and in the same frequency band. In contrast, in the HD mode, the relay receives and transmits in the same frequency band but not at the same time or at the same time but in orthogonal frequency bands, in order to avoid self-interference.
 Given the  limitations of current radio implementations, practical FD relaying suffers from  self-interference. As a result, HD relaying has been widely adopted in the literature \cite{1435648}-\cite{6763001}, \cite{laneman1, laneman2},  \cite{4107956}-\cite{6121995}.

 The capacity of the two-hop FD relay channel  without self-interference has been derived in  \cite{cover} (see the capacity of the degraded relay channel). On the other hand, although extensively investigated, the capacity of the two-hop HD relay channel is not fully known nor understood. The reason for this is that   a capacity  expression which can be easily evaluated is not available and   an explicit coding scheme which achieves the capacity is not known either. Currently, for HD relaying, detailed coding schemes exist only for rates  which are strictly smaller than the capacity,    see \cite{1435648} and \cite{cheap_paper}. To achieve the  rates given in \cite{1435648} and \cite{cheap_paper},   the HD relay   receives a codeword in one time slot, decodes the received codeword, and  re-encodes and re-transmits the decoded information  in the following time slot. However, such fixed switching between reception and  transmission at the HD relay  was shown to be suboptimal in \cite{kramer2004models}. In particular,  in  \cite{kramer2004models},  it  was shown that if the fixed scheduling of reception and transmission at the HD relay   is abandoned, then additional information can be encoded in the relay's reception and transmission switching pattern yielding an increase in data rate. In addition,   it was shown  in \cite{kramer2004models}  that  the HD relay channel can be analyzed using the framework developed  for the  FD relay channel in \cite{cover}. In particular,   results derived for the FD relay channel   in \cite{cover} can be directly applied to the HD relay channel. Thereby, using the converse  for the degraded relay channel in \cite{cover}, the capacity of the discrete memoryless two-hop  HD relay channel is obtained as  \cite{kramer2004models},  \cite{6763001}, \cite{kramer_book_1}
\begin{eqnarray}\label{p-1}
    C=\max_{p(x_1,x_2)}\min\big\{ I(X_1;Y_1|X_2)\;,\; I(X_2;Y_2)\},
\end{eqnarray}
where $I(\cdot;\cdot)$ denotes the mutual information, $X_1$ and $X_2$ are the inputs at source and relay, respectively,  $Y_1$ and $Y_2$ are the outputs at relay and destination, respectively, and $p(x_1,x_2)$ is the joint probability mass function (PMF) of $X_1$ and $X_2$. Moreover, it was shown in \cite{kramer2004models},  \cite{6763001}, \cite{kramer_book_1} that $X_2$ can be represented as $X_2=[X_2',U]$, where $U$ is an auxiliary random variable with two outcomes $t$ and $r$ corresponding to the HD relay transmitting and receiving, respectively. Thereby, (\ref{p-1}) can be written equivalently as
\begin{equation}\label{p-2}
    C=\max_{p(x_1,x_2',u)}\min\big\{ I(X_1;Y_1|X_2',U)\;,\; I(X_2',U;Y_2)\},
\end{equation}
where $p(x_1,x_2',u)$ is the joint PMF of $X_1$, $X_2'$, and $U$.
However, the capacity expressions in (\ref{p-1}) and (\ref{p-2}), respectively, cannot be evaluated since it is not known how $X_1$ and $X_2$  nor $X_1$, $X_2'$, and $U$ are mutually dependent, i.e., $p(x_1,x_2)$ and $p(x_1,x_2',u)$ are not known. In fact, the authors of \cite[page 2552]{6763001} state that: ``\textit{Despite knowing the capacity expression (i.e., expression  (\ref{p-2})), its actual evaluation is elusive as it is not clear what the optimal input distribution $p(x_1,x_2',u)$ is.}'' 
On the other hand, for the  coding scheme that would achieve (\ref{p-1}) and (\ref{p-2}) if $p(x_1,x_2)$ and $p(x_1,x_2',u)$ were known, it can be argued  that it has to be a decode-and-forward strategy since  the two-hop HD relay channel belongs to the class of  the degraded relay channels defined in \cite{cover}.  Thereby,   the HD relay should  decode  the received codewords, map the decoded information to  new codewords, and transmit  them to the destination. Moreover, it is known from  \cite{kramer2004models}  that such a coding scheme requires the HD
relay   to switch between reception and transmission in a symbol-by-symbol manner, and not in a codeword-by-codeword manner as in \cite{1435648} and \cite{cheap_paper}. However, since    $p(x_1,x_2)$  and $p(x_1,x_2',u)$ are not known and since an explicit  coding scheme does not exist, it is currently not known how to evaluate (\ref{p-1}) and (\ref{p-2}) nor how to encode   additional information  in the relay's reception and transmission switching pattern and thereby   achieve (\ref{p-1}) and (\ref{p-2}).

 Motivated by the above discussion, in this paper, 
we derive a new expression for the   capacity of the two-hop HD relay channel by simplifying previously derived converse expressions. In contrast  to previous results,  the new capacity expression can be easily evaluated. Moreover, we propose an explicit coding scheme which     achieves the capacity. 
In particular, we show that achieving the capacity requires the
relay indeed to switch between reception and transmission in a
symbol-by-symbol manner as predicted in \cite{kramer2004models}. Thereby,    the relay does not only send  information   to the destination by   transmitting  information-carrying symbols but also      with the zero symbols resulting from the relay's silence during reception. In addition, we propose a modified coding scheme for practical implementation  where  the HD relay receives and transmits at the same time (i.e., as in FD relaying), however, the simultaneous reception and transmission is performed such that  self-interference is completely avoided.
  As  examples, we compute the capacities of the two-hop HD relay channel for the cases when the source-relay and  relay-destination links are  both binary-symmetric channels (BSCs) and  additive white Gaussian noise (AWGN) channels, respectively, and we numerically  compare  the capacities with the rates achieved by conventional relaying where the relay receives and transmits  in a codeword-by-codeword fashion and switches between  reception and transmission in a strictly alternating manner. Our numerical results  show that the  capacities of the two-hop HD relay channel for BSC and AWGN links are significantly larger than the rates achieved with conventional relaying.


The rest of this paper is organized as follows. In Section~\ref{sec_2}, we present the channel model. In Section~\ref{sec_c-r3}, we derive a new expression for the capacity of the considered channel and propose a corresponding coding scheme. In Section~\ref{sec_c-4}, we investigate the capacity for the cases when the source-relay and relay-destination links are both BSCs and AWGN channels, respectively. In Section~\ref{sec_num}, we numerically evaluate  the derived capacity expressions and compare them to the  rates achieved by conventional relaying. Finally, Section~\ref{sec-conc_c} concludes the paper.

\section{System Model}\label{sec_2}

\begin{figure}
\centering
\includegraphics[width=5in]{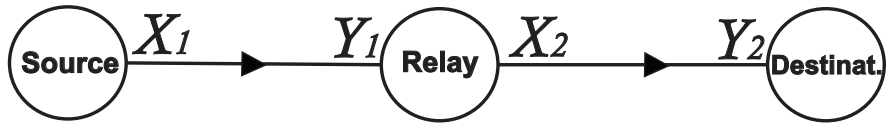}
\caption{Two-hop relay channel.}  \label{fig_1sda}
\end{figure}

The two-hop HD relay channel   consists of a source, an HD relay, and a destination, and the direct link between  source and destination is not available, see Fig.~\ref{fig_1sda}. Due to the HD constraint, the relay cannot transmit and receive at the same time.   In the following, we formally define the channel model.

\subsection{Channel Model} 

The discrete memoryless two-hop HD relay channel is defined by $\mathcal{X}_1$, $\mathcal{X}_2$,     $\mathcal{Y}_1$, $\mathcal{Y}_2$,   and $p(y_1,y_2|x_1,x_2)$, where $\mathcal{X}_1$ and $\mathcal{X}_2$ are the finite  input alphabets at  the source and the relay, respectively,    $\mathcal{Y}_1$ and $\mathcal{Y}_2$ are the finite  output alphabets at   the relay and the destination, respectively,   and $p(y_1,y_2| x_1,x_2)$ is the  PMF  on   $\mathcal{Y}_1\times\mathcal{Y}_2$  for given $x_1\in \mathcal{X}_1$ and $x_2\in \mathcal{X}_2$.  The channel is memoryless in the sense  that given the input symbols   for the $i$-th channel use, the $i$-th output symbols  are independent from all previous input symbols. As a result, the conditional PMF $p(  y_{1}^n, y_{2}^n| x_{1}^n, x_{2}^n)$, where the  notation $a^n$ is   used   to denote the ordered sequence  $a^n=(a_{1}, a_{2},..., a_{n})$, can be factorized as $
     p(  y_{1}^n, y_{2}^n| x_{1}^n, x_{2}^n) =\prod_{i=1}^n p(y_{1i}, y_{2i}| x_{1i}, x_{2i}).$

For the considered channel and the $i$-th channel use, let  $X_{1i}$ and $X_{2i}$ denote the random variables (RVs) which model the input at  source and relay, respectively, and let $Y_{1i}$ and $Y_{2i}$  denote the RVs which model the output at   relay and destination, respectively. 

In the following, we model the HD constraint of the relay and discuss its effect on some important PMFs that will be used throughout this paper.

\subsection{Mathematical Modelling of the HD Constraint}

Due to the HD constraint  of the relay,   the  input  and   output symbols of the relay cannot assume non-zero values at the same time. More precisely, for each channel use, if the input symbol of the relay is non-zero then the output symbol has to be zero, and vice versa, if the output symbol of  the relay is non-zero then the input symbol has to be zero. Hence, the following holds
\begin{equation}\label{eq_Y_rv}
     Y_{1i}=\left\{ \hspace{-2mm}
\begin{array}{ll}
Y_{1i}' &  \hspace{-1mm} \textrm{if } X_{2i}=0 \\
0 &  \hspace{-1mm} \textrm{if } X_{2i}\neq 0 ,
\end{array} \right.
\end{equation}
where  $Y_{1i}'$ is an RV that take values from  set  $\mathcal{Y}_{1}$.

In order to model the HD constraint of the relay more conveniently, we represent the input set of the relay  $\mathcal{X}_2$ as the union of two sets $\mathcal{X}_2=\mathcal{X}_{2R}\cup \mathcal{X}_{2T}$, where $\mathcal{X}_{2R}$ contains only one element, the zero symbol, and $\mathcal{X}_{2T}$ contains all symbols in $\mathcal{X}_2$ except the zero symbol. Note that, because of the HD constraint, $\mathcal{X}_2$ has to contain the zero symbol, i.e., the relay has to be silent in some portion of the time during which the relay can receive.
Furthermore,   we  introduce an auxiliary random variable, denoted by $U_i$, which takes values from the set $\{t,r\}$, where $t$ and $r$ correspond to the relay transmitting a non-zero symbol and a zero symbol, respectively. Hence, $U_i$ is defined as
\begin{equation}\label{eq_U_i}
     U_i=\left\{ \hspace{-2mm}
\begin{array}{ll}
r &  \hspace{-1mm} \textrm{if } X_{2i}=0 \\
t &  \hspace{-1mm} \textrm{if } X_{2i}\neq 0 .
\end{array} \right.
\end{equation}
Let us denote  the probabilities of the relay transmitting a non-zero and a zero symbol for the $i$-th channel use as ${\rm Pr}\{U_i=t\}={\rm Pr}\{X_{2i}\neq 0\}= P_{U_i}$ and ${\rm Pr}\{U_i=r\}={\rm Pr}\{X_{2i}= 0\}=1-P_{U_i}$, respectively.  
We now use (\ref{eq_U_i}) and represent  $X_{2i}$    as a  function  of the outcome of $U_i$. Hence, we have
\begin{eqnarray}\label{eq_X_2i}
     X_{2i}=\hspace{-1mm}\left\{ \hspace{-2mm}
\begin{array}{ll}
0 & \hspace{-2mm} \textrm{if } U_i=r \\
V_{i} & \hspace{-2mm}\textrm{if } U_i=t ,
\end{array}\hspace{-1.5mm} \right. 
\end{eqnarray}
where $V_{i}$ is an RV with distribution $p_{V_i}(x_{2i})$ that takes values from the set  $\mathcal{X}_{2T}$, or   equivalently, an    RV which takes values from the set  $\mathcal{X}_{2}$, but with $p_{V_i}(x_{2i}=0)=0$.
From (\ref{eq_X_2i}), we obtain  
\begin{align}
p(x_{2i}|U_i=r)&=\delta(x_{2i}),\label{eq_rt1}\\
p(x_{2i}|U_i=t)&=p_{V_i}(x_{2i}),\label{eq_rt1a}
\end{align}
where $\delta(x)=1$ if $x=0$ and $\delta(x)=0$ if $x\neq 0$. Furthermore, for the derivation of the capacity,
  we will also need the conditional PMF  $p(x_{1i}|x_{2i}=0)$  which is the  input distribution at the source when  the relay transmits a zero (i.e., when $U_i=r$). As we will see in Theorem~\ref{SItheo_1}, the  distributions  $p(x_{1i}|x_{2i}=0)$ and $p_{V_i}(x_{2i})$ have to be optimized in order to achieve the capacity. 
 Using $p(x_{2i}|U_i=r)$ and $p(x_{2i}|U_i=t)$, and the law of total probability, the PMF of $X_{2i}$, $p(x_{2i})$, is obtained as
\begin{eqnarray}\label{eq_p(x_2)}
    p(x_{2i}) &=&p(x_{2i}|U_i=t) P_{U_i}+ p(x_{2i}|U_i=r)(1-P_{U_i})\nonumber\\
&\stackrel{(a)}{=} &p_{V_i}(x_{2i}) P_{U_i}+ \delta(x_{2i})(1-P_{U_i}),
\end{eqnarray}
where   $(a)$ follows from (\ref{eq_rt1}) and (\ref{eq_rt1a}). In addition, we will  also need the distribution of $Y_{2i}$, $p(y_{2i})$, which, using the  law of total probability,   can be written  as
\begin{eqnarray}\label{eqe_1}
    p(y_{2i})=  p(y_{2i}|U_i=t) P_{U_i}+ p(y_{2i}|U_i=r) (1-P_{U_i}).
\end{eqnarray}
On the other hand, using $X_{2i}$ and the  law of total probability, $p(y_{2i}|U_i=r)$ can be written as
\begin{align}
p(y_{2i}|U_i=r) &= \sum_{x_{2i}\in \mathcal{X}_2}p(y_{2i},x_{2i}|U_i=r) = \sum_{x_{2i}\in \mathcal{X}_2}p(y_{2i}|x_{2i}, U_i=r) p(x_{2i}|U_i=r)\nonumber\\
 &\stackrel{(a)}{=}   
\sum_{x_{2i}\in \mathcal{X}_2} p(y_{2i}|x_{2i}, U_i=r) \delta(x_{2i}) =  
p(y_{2i}|x_{2i}=0, U_i=r)\nonumber\\ 
 &\stackrel{(b)}{=}  p(y_{2i}|x_{2i}=0)   \label{eq_py2_r},
\end{align}
where $(a)$ is due to (\ref{eq_rt1}) and   $(b)$ is the result of conditioning on the same variable twice since if $X_{2i}=0$ then $U_i=r$, and vice versa.  
On the other hand, using $X_{2i}$ and the law of total probability, $p(y_{2i}|U_i=t)$ can be written as
\begin{align}
   p(y_{2i}|U_i=t)& = \sum_{x_{2i}\in \mathcal{X}_2}p(y_{2i},x_{2i}|U_i=t) = \sum_{x_{2i}\in\mathcal{X}_{2}}  p(y_{2i}|x_{2i}, U_i=t) p(x_{2i}|U_i=t) \nonumber\\
& \stackrel{(a)}{=}  \sum_{x_{2i}\in\mathcal{X}_{2T}}  p(y_{2i}|x_{2i}, U_i=t) p_{V_i}(x_{2i}) \stackrel{(b)}{=}  \sum_{x_{2i}\in\mathcal{X}_{2T}}  p(y_{2i}|x_{2i}) p_{V_i}(x_{2i}) 
\label{eq_py2_t},
\end{align}
where $(a)$ follows from (\ref{eq_rt1a}) and since $V_{i}$  takes values from  set $\mathcal{X}_{2T}$, and $(b)$ follows since conditioned on $X_{2i}$, $Y_{2i}$ is independent of $U_i$. In (\ref{eq_py2_t}),  $p(y_{2i}|x_{2i})$ is the distribution at the output of the  relay-destination channel conditioned on  the relay's input  $X_{2i}$.

\subsection{Mutual Information and Entropy}\label{sec-H(Y_2)}

For the capacity expression given later in Theorem~\ref{SItheo_1}, we need  $I(X_1;Y_1|X_2=0)$, which is the mutual information between the source's input $X_1$ and the relay's output $Y_1$ conditioned on the relay  having its input  set to $X_2=0$, and $I(X_2;Y_2)$, which is the mutual information between the relay's input $X_2$ and the destination's output $Y_2$.

The mutual information $I(X_1;Y_1|X_2=0)$ is obtained by definition as
\begin{align}\label{er_ksndk}
 I\big(X_{1}; Y_{1}|& X_2=0 \big) =\sum_{x_1\in \mathcal{X}_1} \sum_{y_1\in \mathcal{Y}_1} p(y_1|x_1,x_2=0 )   p(x_1| x_2=0 ) \log_2\left( \frac{p(y_1|x_1,x_2=0 )}{p(y_1|x_2=0 )} \right),
\end{align}
where   
\begin{align}\label{sdswe}
p(y_1|x_2=0)=\sum_{x_1\in \mathcal{X}_1} p(y_1|x_1,x_2=0 ) p(x_1|x_2=0) .
\end{align}
In (\ref{er_ksndk}) and (\ref{sdswe}), $p(y_1|x_1,x_2=0)$ is the distribution at the output of the source-relay channel conditioned on  the relay  having its input  set to $X_2=0$, and conditioned  on  the input symbols at the source  $X_1$.  

On the other hand,  $I(X_2;Y_2)$ is given by
\begin{equation}\label{eq_ident}
    I(X_2;Y_2)=H(Y_2)-H(Y_2|X_2),
\end{equation}
where    $H(Y_2)$ is the entropy of RV $Y_2$, and $H(Y_2|X_2)$ is the   entropy of $Y_2$ conditioned on   $X_2$. The entropy $H(Y_2)$ can be found by definition as 
\begin{align}\label{eq_H(y2)dad}
      H(Y_2) &=-\sum_{y_{2}\in\mathcal{Y}_2}  p(y_{2}) \log_2(p(y_{2})) \nonumber\\
& \stackrel{(a)}{=} -\sum_{y_{2}\in\mathcal{Y}_2} \big[ p(y_{2}|U=t) P_{U}+ p(y_{2}|U=r) (1-P_{U}) \big] \nonumber\\
&\qquad \times\log_2\big[ p(y_{2}|U=t) P_{U}+ p(y_{2}|U=r) (1-P_{U})\big], \qquad
\end{align}
where $(a)$ follows from (\ref{eqe_1}). Now, inserting $p(y_{2}|U=r)$ and $p(y_{2}|U=t)$ given in (\ref{eq_py2_r}) and (\ref{eq_py2_t}), respectively, into (\ref{eq_H(y2)dad}), we obtain the final expression for $H(Y_2)$, as 
\begin{align}\label{eq_H(y2)}
    H(Y_2) &= -\sum_{y_{2}\in\mathcal{Y}_2} \bigg[ P_{U} \sum_{x_{2}\in\mathcal{X}_{2T}}  p(y_{2}|x_{2}) p_{V}(x_{2})  + p(y_{2}|x_{2}=0) (1-P_{U}) \bigg] \nonumber\\
&\qquad\times \log_2\bigg[ P_{U} \hspace{-3mm}\sum_{x_{2}\in\mathcal{X}_{2T}}   \hspace{-3mm} p(y_{2}|x_{2})p_{V}(x_{2}) + p(y_{2}|x_{2}=0) (1-P_{U}) \bigg]. 
\end{align}
On the other hand, the conditional  entropy $H(Y_2|X_2)$ can be found  based on its definition as
\begin{align}\label{eq_H(y2|x2)}
     H(Y_2|X_2)& =-\sum_{x_{2}\in\mathcal{X}_2} p(x_2) \sum_{y_{2}\in\mathcal{Y}_2}  p(y_{2}|x_{2}) \log_2(p(y_{2}|x_2))\nonumber\\
 &\stackrel{(a)}{=}-P_U\sum_{x_{2}\in\mathcal{X}_{2T}} p_{V}(x_{2}) \sum_{y_{2}\in\mathcal{Y}_2}  p(y_{2}|x_{2}) \log_2(p(y_{2}|x_2))\nonumber\\
&\quad \;- (1-P_U) \sum_{y_{2}\in\mathcal{Y}_2}  p(y_{2}|x_{2}=0) \log_2(p(y_{2}|x_2=0)), 
\end{align}
where $(a)$ follows by inserting $p(x_2)$  given in (\ref{eq_p(x_2)}). 
Inserting   $H(Y_2)$ and $H(Y_2|X_2)$ given in (\ref{eq_H(y2)}) and (\ref{eq_H(y2|x2)}), respectively, into (\ref{eq_ident}), we obtain the final expression for $I(X_2;Y_2)$, which is dependent on $p(x_2)$, i.e., on $p_{V}(x_{2})$ and $P_U$. To highlight this dependence, we sometimes write $I(X_2;Y_2)$ as  $I(X_2;Y_2)\Big|_{P_U}$.

We are now ready to present a new  capacity expression for the considered relay channel.

\section{New Capacity Expression and Explicit Coding Scheme}\label{sec_c-r3}

In this section, we provide a new and   easy-to-evaluate expression for the capacity of the two-hop HD relay channel by  simplifying previously derived converse expressions and provide an explicit coding scheme  which achieves the capacity.

\subsection{New Capacity Expression}\label{sec_c-3}

A new expression for the capacity  of the two-hop HD relay channel is given in the following theorem.

\begin{theorem}\label{SItheo_1}
The capacity of the two-hop HD  relay channel  is given by
\begin{align}\label{SIeq_capacity_2}
    C  =\max_{P_U}\min\left\{   \max_{p(x_{1}|x_2=0)} \hspace{-0.5mm} I\big(X_{1}; Y_{1}| X_2=0\big)(1-P_{U}),   \max_{p_{V}(x_{2})}   I(X_2; Y_{2})\big|_{P_{U} }  \right\},
\end{align}
where $I\big(X_{1}; Y_{1}| X_2=0 \big)$ is given in (\ref{er_ksndk}) and  $I(X_2; Y_{2})$ is given in (\ref{eq_ident})-(\ref{eq_H(y2|x2)}). The optimal $P_U$ that maximizes the capacity in  (\ref{SIeq_capacity_2}) is given by $P_U^*=\min\{P_U',P_U''\}$, where $P_U'$ is the solution  of
\begin{align}\label{eq_PU'}
   \max_{p(x_{1}|x_2=0)} \hspace{-0.5mm} I\big(X_{1}; Y_{1}| X_2=0\big)(1-P_{U})=  \max_{p_{V}(x_{2})}   I(X_2; Y_{2})\big|_{P_{U} } ,  
\end{align}
where, if (\ref{eq_PU'}) has two solutions, then $P_U'$ is the smaller  of the two, 
and $P_U''$ is the solution of
\begin{align}\label{eq_PU''}
  \frac{\partial  \Big(\max\limits_{p_{V}(x_{2})}   I(X_2; Y_{2})\big|_{P_{U} } \Big)}{\partial P_U}=0.
\end{align}
 If $P_U^*=P_U'$,  the capacity   in  (\ref{SIeq_capacity_2})  simplifies to
\begin{align}\label{SIeq_capacity_2b}
    C  =   \max_{p(x_{1}|x_2=0)} \hspace{-0.5mm} I\big(X_{1}; Y_{1}| X_2=0\big)(1-P_{U}') =   \max_{p_{V}(x_{2})}   I(X_2; Y_{2})\big|_{P_{U}=P_U' } ,
\end{align}
 whereas, if $P_U^*=P_U''$,  the capacity   in  (\ref{SIeq_capacity_2}) simplifies to
\begin{align}\label{SIeq_capacity_2a}
    C  =    \max_{p_{V}(x_{2})}   I(X_2; Y_{2})\big|_{P_{U}=P_U'' }    = \max_{p(x_{2})}   I(X_2; Y_{2}) ,
\end{align}
which is the capacity of the relay-destination channel.
\end{theorem}
\begin{IEEEproof}
To derive the new capacity expression in (\ref{SIeq_capacity_2}), we combine the results from \cite{cover} and \cite{kramer2004models}. In particular,   \cite{kramer2004models} showed that the HD relay channel can be analyzed with the framework  developed for the FD relay channel in \cite{cover}. Since the considered two-hop HD relay channel belongs to the class of degraded relay channels  defined in \cite{cover}, the rate of this channel is upper bounded by  \cite{cover}, \cite{kramer2004models}
\begin{eqnarray}\label{eq_1_0a1}
 &&\hspace{-7mm}    R\leq  \max_{p(x_1,x_2)}  \min \big\{  I\big(X_{1}; Y_{1}|X_{2}\big) ,    I\big(X_{2};Y_{2}\big)\big\}  .
\end{eqnarray}
On the other hand, $I\big(X_{1}; Y_{1}|X_{2}\big)$ can be simplified as
\begin{align}\label{rav_1aa}
  I\big(X_{1}; Y_{1}|X_2\big) 
  &  =   I\big(X_{1}; Y_{1}| X_{2}=0\big) (1-P_{U}) +   I\big(X_{1}; Y_{1}|X_{2}\neq 0 \big) P_{U} \nonumber\\
&  
 \stackrel{(a)}{=}
   I\big(X_{1}; Y_{1}| X_{2}=0\big) (1-P_{U}),  
\end{align}
where    $(a)$ follows from (\ref{eq_Y_rv}) since  when $X_{2}\neq 0$,  $Y_{1}$ is deterministically zero thereby leading to $I\big(X_{1}; Y_{1}|X_{2}\neq 0\big)= 0$. 
 Inserting  (\ref{rav_1aa}) into (\ref{eq_1_0a1}),  we obtain
\begin{equation}\label{eq_1_0a2}
    R\leq  \max_{p(x_1,x_2)}  \min \big\{  I\big(X_{1}; Y_{1}| X_{2}=0\big) (1-P_{U})   \; ,\;    I\big(X_{2};Y_{2}\big)\big\}  .
\end{equation}
Note that the PMF $p(x_1,x_2)$ can be  written equivalently as
\begin{align}\label{eq_c2}
p(x_1,x_2)=p(x_1|x_2)p(x_2)  \stackrel{(a)}{=}  p(x_1|x_2) \big( p_V(x_2) P_U+ \delta(x_2) (1-P_U)\big),
\end{align}
where $(a)$ follows from (\ref{eq_p(x_2)}).
As a result, the maximization over $p(x_1,x_2)$ can be resolved into a joint maximization over  $p(x_1|x_2)$,  $p_V(x_2)$, and $P_U$. Thereby, (\ref{eq_1_0a2})  can be written equivalently as
\begin{align}\label{eq_c3}
 R\leq \max_{p(x_1|x_2),\; p_V(x_2),\; P_U}   \min \big\{  I\big(X_{1}; Y_{1}| X_{2}=0\big) (1-P_{U})   \; ,\;    I\big(X_{2};Y_{2}\big)\big\} . 
\end{align}
Now, note that  $I\big(X_{1}; Y_{1}| X_{2}=0\big)$ and $I\big(X_{2};Y_{2}\big)$ are dependent only on $p(x_1|x_2=0)$ and $p_V(x_2)$, respectively, and no  other function in the right hand side of (\ref{eq_c3}) is dependent on $p(x_1|x_2=0)$ and $p_V(x_2)$. Therefore, (\ref{eq_c3}) can be written equivalently as
\begin{align}\label{eq_1_0a}
 R\leq \max_{ P_U}   \min \left\{ \max_{p(x_1|x_2=0)}   I\big(X_{1}; Y_{1}| X_{2}=0\big) (1-P_{U})   \; ,\;  \max_{ p_V(x_2)}   I\big(X_{2};Y_{2}\big)  \Big|_{P_U}\right\},
\end{align}
where    the mutual informations $ I\big(X_{1}; Y_{1}| X_{2}=0\big)$ and  $I\big(X_{2};Y_{2}\big)\big|_{P_U}$ are concave functions with respect to $p(x_1|x_2=0)$ and $p_{V}(x_{2})$, respectively\footnote{For concavity of  $ I\big(X_{1}; Y_{1}| X_{2}=0\big)$ and  $I\big(X_{2};Y_{2}\big)\big|_{P_U}$  with respect to $p(x_1|x_2=0)$ and $p(x_{2})$, respectively, see \cite{cover2012elements}. On the other hand, since $1-P_U$ is just the probability $p(x_2=0)$ and  since $p_V(x_2)$ contains the rest of the probability constrained parameters  in $p(x_2)$,  $I(X_2; Y_{2})$ is a jointly concave  function with respect to $p_{V}(x_{2})$ and $P_U$.}. 
Now, note that the right hand side of the expression in (\ref{eq_1_0a}) is identical to the capacity expression in (\ref{SIeq_capacity_2}).   
 The rest of the theorem follows from solving (\ref{SIeq_capacity_2}) with respect to $P_U$, and simplifying the result. In particular, note that the first  term inside the  $\min\{\cdot\}$ function in (\ref{SIeq_capacity_2}) is a decreasing   function  with respect to $P_U$. This function achieves its maximum for $P_U=0$ and its minimum, which is zero, for $P_U=1$. On the other hand, the second  term inside the  $\min\{\cdot\}$ function in (\ref{SIeq_capacity_2}) is a concave   function  with respect to $P_U$\footnote{In \cite[pp. 87-88]{Boyd_CO}, it is proven that if $f(x,y)$ is a jointly concave function in both $(x,y)$ and $\mathcal{C}$ is a convex nonempty set, then the function $g(x)=\underset{y\in\mathcal{C}}{\max} f(x,y)$ is concave in $x$. Using this result, and noting that $I(X_2; Y_{2})$ is a jointly concave  function with respect to $p_{V}(x_{2})$ and $P_U$, we can   conclude that $\underset{p_V(x_2)}{\max} \, I(X_2,Y_2)\big|_{P_U}$ is concave with respect to $P_U$.}. Moreover, this function  is zero when $P_U=0$, i.e., when the relay is always silent.
Now, the maximization of the minimum  of the decreasing and concave functions with respect to $P_U$, given in (\ref{SIeq_capacity_2}),   has a solution $P_U=P_U''$, when the concave function reaches its maximum, found from (\ref{eq_PU''}), and when for this point, i.e., for  $P_U=P_U''$, the decreasing function is larger than the concave function. Otherwise, the solution is $P_U=P_U'$ which  is found from (\ref{eq_PU'}) and in which case $P_U'<P_U''$ holds. If (\ref{eq_PU'})  has two solutions, then $P_U'$ has to be the smaller of the two since $ \max\limits_{p(x_{1}|x_2=0)} \hspace{-0.5mm} I\big(X_{1}; Y_{1}| X_2=0\big)(1-P_{U})$   is a decreasing function with respect to $P_U$. Now, when $P_U^*=P_U'$, (\ref{eq_PU'}) holds, (\ref{SIeq_capacity_2}) simplifies to (\ref{SIeq_capacity_2b}). Whereas, when $P_U^*=P_U''$, then $\max\limits_{p_{V}(x_{2})}   I(X_2; Y_{2})\big|_{P_{U}=P_U'' } = \max\limits_{p_{V}(x_{2})} \max\limits_{P_U} I(X_2; Y_{2}) = \max\limits_{p(x_{2})}  I(X_2; Y_{2})$, thereby leading to  (\ref{SIeq_capacity_2a}). This concludes the proof.
\end{IEEEproof}
%

\subsection{Explicit Capacity Achieving Coding Scheme}\label{sec_ach}
Since an explicit capacity coding scheme is not available in the literature, in the following, we propose an explicit coding scheme which achieves the capacity in (\ref{SIeq_capacity_2}).

\begin{figure}
\centering
 \resizebox{1\linewidth}{!}{
\pstool[width=1\linewidth]{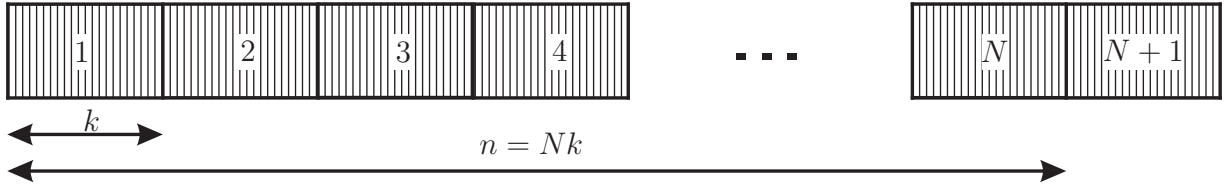}{
\psfrag{1}[c][c][1]{$1$}
\psfrag{2}[c][c][1]{$2$}
\psfrag{3}[c][c][1]{$3$}
\psfrag{4}[c][c][1]{$4$}
\psfrag{N}[c][c][1]{$N$}
\psfrag{N+1}[c][c][1]{$N+1$}
\psfrag{k}[c][c][1]{$k$}
\psfrag{n}[c][c][1]{$n=Nk$}
}}
\caption{To achieve the capacity in (\ref{SIeq_capacity_2}), transmission is organized in $N+1$ blocks and each block comprises $k$ channel uses.}
\label{Fig:Blocks}
\end{figure}

In the following, we describe a method for transferring $n R$ bits of information in $n+k$ channel uses,  where $n,k\to\infty$   and   $n/(n+k)\to 1$ as $n,k\to\infty$. As a result, the information is transferred at rate $R$.  To this end, the transmission is carried out in   $N+1$ blocks, where $N\to\infty$. In each block, we use the channel $k$ times. The numbers  $N$ and $k$ are chosen such that $n=Nk$ holds.   The transmission in $N+1$ blocks is illustrated in Fig.~\ref{Fig:Blocks}.

The source  transmits  message $W$, drawn uniformly from  message set $\{1,2,...,2^{nR}\}$, from the source via the HD relay to the destination. To this end, before the start of   transmission,   message $W$ is spilt into $N$ messages, denoted by $w(1),...,w(N)$, where each $w(i)$, $\forall i$, contains $kR$ bits of information.   The transmission is carried out in the following manner. In   block one,   the source sends message  $w(1)$   in $k$ channel uses to the relay and the relay is silent.  In  block $i$, for $i=2,...,N$, source and relay send  messages $w(i)$ and $w(i-1)$ to relay and destination, respectively,   in $k$ channel uses. In block $N+1$, the relay sends  message  $w(N)$   in $k$ channel uses to the destination and the source is silent. Hence, in the first block and in the $(N+1)$-th block, the relay and the source are silent, respectively, since in the first block the relay does not have information to transmit, and in block $N+1$, the source   has no more information to transmit. In blocks $2$ to $N$, both   source and relay transmit, while meeting the HD constraint in every channel use.  Hence, during the $N+1$ blocks, the channel is used $k(N+1)$ times to send $nR=NkR$ bits of information, leading to an   overall information rate   given by
\begin{eqnarray}
  \lim_{N\to\infty} \lim_{k\to\infty}  \frac{ Nk R}{k(N+1)}=R \;\;\textrm{ bits/use}.
\end{eqnarray} 

A detailed description of the proposed coding scheme is given in  the following, where we explain the  rates, codebooks, encoding, and decoding  used for  transmission.

\textit{Rates: } The transmission rate  of  both source and relay is denoted by $R$ and  given by 
\begin{equation}
   R=C    -\epsilon ,\label{SIeq_r_2} 
\end{equation}
 where  $C$ is given in Theorem 1  and $\epsilon>0$ is an arbitrarily small number. 
Note that $R$ is  a function of $P_U^*$, see  Theorem~\ref{SItheo_1}.

\textit{Codebooks: }
We have two codebooks: The source's transmission  codebook and the relay's transmission  codebook.

The source's transmission codebook is generated by mapping each possible binary sequence comprised of $k R$ bits, where $R$ is given by (\ref{SIeq_r_2}),  to a codeword\footnote{The subscript $1|r$ in $\mathbf{x}_{1|r}$  is used to indicate that   codeword $\mathbf{x}_{1|r}$ is comprised of symbols which are  transmitted by the source only when   $U_i=r$, i.e., when $X_{2i}=0$.}  $\mathbf{x}_{1|r}$   comprised of $k (1-P_U^*)$  symbols. The symbols in each codeword  $\mathbf{x}_{1|r}$  are generated independently according to   distribution  $p(x_{1}|x_2=0)$. Since in total  there are $2^{k R}$ possible binary sequences comprised of  $k R $  bits, with this mapping we generate $2^{k R }$  codewords $\mathbf{x}_{1|r}$ each  containing $k (1-P_U^*)$ symbols.  These $2^{k R }$  codewords  form the source's transmission codebook, which we  denote    by $\mathcal{C}_{1|r}$.

The relay's transmission  codebook is generated by mapping each possible binary sequence comprised of $k R $ bits, where $R $ is given by (\ref{SIeq_r_2}),  to a  transmission codeword  $\mathbf{x}_2$   comprised of $k$  symbols. The $i$-th symbol, $i=1,...,k$, in  codeword   $\mathbf{x}_2$ is generated in the following manner. For each symbol a coin is tossed. The coin is such that it produces   symbol $r$ with probability $1-P_U^*$ and   symbol $t$ with probability $P_U^*$. If the outcome of the coin flip is $r$, then the  $i$-th symbol of the relay's transmission codeword $\mathbf{x}_2$ is set to zero. Otherwise, if the outcome of the coin flip is $t$, then the $i$-th symbol of  codeword  $\mathbf{x}_2$  is generated independently according to   distribution  $p_{V}(x_{2})$. In this way, the symbols in  $\mathbf{x}_2$ are distributed according to the distribution $p(x_2)$ given in (\ref{eq_p(x_2)}).  The  $2^{k R}$  codewords $\mathbf{x}_2$   form the relay's transmission codebook denoted  by $\mathcal{C}_2$.

The two codebooks are known at all three nodes.

\textit{Encoding,  Transmission, and Decoding: }
In the first block, the source   maps $w(1)$  to the appropriate codeword $\mathbf{x}_{1|r}(1)$ from its codebook $\mathcal{C}_{1|r}$. Then, codeword $\mathbf{x}_{1|r}(1)$ is transmitted  to the relay, which is scheduled to always  receive and be silent (i.e., to set its input to zero)  during the  first block. However,  knowing that the transmitted codeword from the source $\mathbf{x}_{1|r}(1)$ is comprised of $k (1-P_U^*)$ symbols, the relay constructs the received codeword, denoted by $\mathbf{y}_{1|r}(1)$, only from the first  $k (1-P_U^*)$  received symbols.   
In Appendix~\ref{app_proof_of_prob_error_1} , we prove that  codeword $\mathbf{x}_{1|r}(1)$ sent in the first block can be decoded successfully from the received codeword at the relay $\mathbf{y}_{1|r}(1)$ using a typical decoder \cite{cover2012elements} since $R $ satisfies
\begin{eqnarray}\label{eq_d_1aa}
   R  < \max_{p(x_{1}|x_2=0)} I\big(X_{1}; Y_{1}| X_2=0\big) (1-P_U^*).
\end{eqnarray}

In blocks $i=2,...,N$, the encoding,  transmission, and decoding are performed as follows. In    blocks $i=2,...,N$, the source and the relay map $w(i)$ and $w(i-1)$ to the appropriate codewords $\mathbf{x}_{1|r}(i)$ and $\mathbf{x}_2(i)$ from codebooks $\mathcal{C}_{1|r}$ and $\mathcal{C}_{2}$, respectively. Note that the source also knows  $\mathbf{x}_2(i)$ since  $\mathbf{x}_2(i)$ was generated from $w(i-1)$ which the source transmitted in the previous (i.e., $(i-1)$-th)   block.
The transmission of  $\mathbf{x}_{1|r}(i)$ and $\mathbf{x}_2(i)$ can be performed in two ways: 1) by the relay  switching between reception and transmission, and 2) by the relay always receiving and transmitting as in FD relaying. We first explain the first option.

Note that both  source and   relay know the position of the zero symbols in $\mathbf{x}_2(i)$. Hence, if the  first symbol  in  codeword $\mathbf{x}_2(i)$ is zero, then in the first symbol interval of block $i$, the source transmits its first symbol from codeword  $\mathbf{x}_{1|r}(i)$ and the relay receives. By receiving, the relay actually also sends the first symbol of codeword $\mathbf{x}_2(i)$, which is the symbol zero, i.e., $x_{21}=0$.   On the other hand, if the first symbol  in   codeword $\mathbf{x}_2(i)$  is non-zero, then in the first symbol interval of block $i$, the relay transmits its first symbol from codeword  $\mathbf{x}_2(i)$ and the source is silent. 
 The same procedure  is performed   for the $j$-th channel use in block $i$, for $j=1,...,k$. In particular,  if the  $j$-th symbol  in  codeword $\mathbf{x}_2(i)$ is zero, then in the $j$-th channel use of block $i$  the source transmits its next untransmitted symbol from codeword  $\mathbf{x}_{1|r}(i)$ and the relay receives. With this reception, the relay actually also sends the $j$-th symbol of codeword $\mathbf{x}_2(i)$, which is the symbol zero, i.e., $x_{2j}=0$. On the other hand, if the $j$-th symbol  in   codeword $\mathbf{x}_2(i)$ is non-zero, then for the $j$-th channel use of block $i$, the relay transmits the $j$-th symbol of codeword  $\mathbf{x}_2(i)$ and the source is silent. Note that codeword $\mathbf{x}_2(i)$ contains   $k(1-P_U^*)\pm \varepsilon(i)$ symbols zeros, where $\varepsilon(i)>0$. Due to the strong law of large numbers \cite{cover2012elements}, $\lim\limits_{k\to\infty} \varepsilon(i)/k=0$ holds, which means that for large  enough $k$, the fraction of symbols zeros in codeword $\mathbf{x}_2(i)$ is $1-P_U^*$. Hence,    for $k\to\infty$, the source   can transmit practically all\footnote{When we say  practically all, we mean either all or all except for a negligible fraction $\lim_{k\to\infty} \varepsilon(i)/k=0$ of them.} of its $k(1-P_U^*)$ symbols from codeword $\mathbf{x}_{1|r}(i)$ during a single block to the relay. Let  $\mathbf{y}_{1|r}(i)$ denote the corresponding received   codeword at the relay.   In  Appendix~\ref{app_proof_of_prob_error_1}, we prove that the codewords $\mathbf{x}_{1|r}(i)$ sent in blocks $i=2,\dots,N$ can be decoded successfully at the relay from the corresponding received codewords $\mathbf{y}_{1|r}(i)$  using a typical decoder \cite{cover2012elements} since   $R$ satisfies (\ref{eq_d_1aa}). Moreover, in  Appendix~\ref{app_proof_of_prob_error_1}, we also prove that, for $k\to\infty$,   the codewords $\mathbf{x}_{1|r}(i)$  can be  successfully decoded at the relay  even though, for some blocks $i=2,...,N$, only $k(1-P_U^*)-\varepsilon(i)$ symbols out of $k(1-P_U^*)$ symbols in  codewords  $\mathbf{x}_{1|r}(i)$  are transmitted to the relay.
On the other hand, the relay  sends  the entire codeword $\mathbf{x}_2(i)$, comprised of  $k$ symbols of which a fraction $1-P_U^*$ are zeros,  to the destination. In particular, the relay sends the  zero symbols of codeword $\mathbf{x}_2(i)$ to the destination by being silent during reception, and sends the non-zero  symbols of codeword $\mathbf{x}_2(i)$ to the destination by actually transmitting them.   On  the other hand, the destination   listens during the entire block and   receives a codeword  $\mathbf{y}_2(i)$.   By    following  the ``standard" method in \cite[Sec.~7.7]{cover2012elements} for analyzing the probability of error for rates smaller than the capacity, it can be shown in a straightforward manner that the destination  can successfully decode  $\mathbf{x}_2(i)$ from the received codeword $\mathbf{y}_2(i)$, and thereby obtain $w(i-1)$, since   rate $R$ satisfies  
\begin{eqnarray}
     R <\max_{p_{V}(x_{2})} I(X_2; Y_{2})\Big|_{P_{U}=P_{U}^* }  .     \label{SIeq_r_2a}
\end{eqnarray}

In a practical implementation,  the relay may not be able to switch between reception and transmission in a symbol-by-symbol manner, due to  practical constraints   regarding the speed of switching. Instead, we may allow the relay to receive and transmit at the same time and in the same frequency band similar to FD relaying. However, this simultaneous reception and transmission is performed while  avoiding self-interference  since, in each symbol interval,  either the  input or the  output information-carrying symbol   of the relay is zero. 
This is accomplished   in the following manner. The source performs the same operations  as for the case   when the relay switches between reception and transmission. On the other hand, the relay transmits all symbols from $\mathbf{x}_2(i)$ while continuously listening. Then, the relay discards from the received codeword, denoted by  $\mathbf{y}_1(i)$,   those symbols for which the corresponding symbols in $\mathbf{x}_2(i)$ are non-zero, and only collects the symbols in $\mathbf{y}_1(i)$ for which the corresponding symbols in    $\mathbf{x}_2(i)$ are equal to zero. The collected symbols from $\mathbf{y}_1(i)$ constitute the relay's   information-carrying received   codeword $\mathbf{y}_{1|r}(i)$ which is used for decoding. Codeword $\mathbf{y}_{1|r}(i)$  is completely free of self-interference since the symbols in $\mathbf{y}_{1|r}(i)$  were received in symbol intervals for which the corresponding transmit symbol at the relay was zero.

In the last (i.e., the $(N+1)$-th) block, the source is silent and the relay transmits $w(N)$ by mapping it to the corresponding   codeword   $\mathbf{x}_2(i)$ from set $\mathcal{C}_{2}$. The relay transmits all   symbols in codeword   $\mathbf{x}_2(i)$ to the destination. The destination  can decode the received codeword in block $  N+1 $ successfully, since (\ref{SIeq_r_2a}) holds.

Finally, since both  relay and destination can decode their respective codewords  in each block, the entire message $W$ can be decoded successfully at the destination at the end of the $(N+1)$-th block.

\subsubsection{Coding Example}

\begin{figure}
\centering
\resizebox{1\linewidth}{!}{
\pstool[width=1\linewidth]{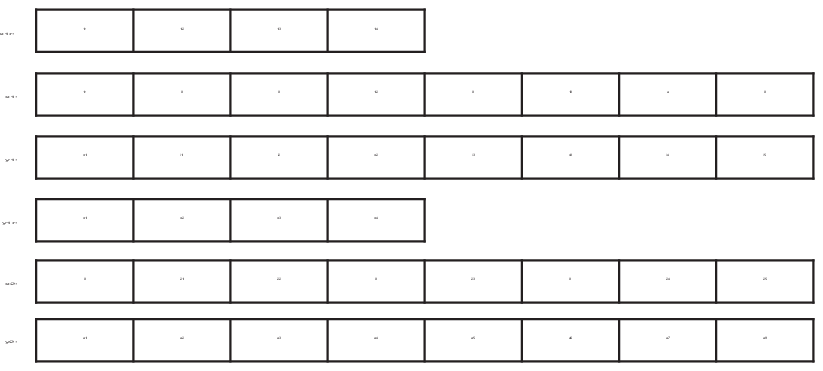}{
\psfrag{21}[c][c][1]{$x_{22}$}
\psfrag{22}[c][c][1]{$x_{23}$}
\psfrag{23}[c][c][1]{$x_{25}$}
\psfrag{24}[c][c][1]{$0$}
\psfrag{25}[c][c][1]{$x_{28}$}
\psfrag{1:}[c][c][1]{$x_{11|r}$}
\psfrag{12}[c][c][1]{$x_{12|r}$}
\psfrag{13}[c][c][1]{$x_{13|r}$}
\psfrag{14}[c][c][1]{$x_{14|r}$}
\psfrag{o1}[c][c][1]{$y_{11}$}
\psfrag{o2}[c][c][1]{$y_{14}$}
\psfrag{o3}[c][c][1]{$y_{16}$}
\psfrag{o4}[c][c][1]{$y_{17}$}
\psfrag{i1}[c][c][1]{$y_{12}$}
\psfrag{i2}[c][c][1]{$y_{13}$}
\psfrag{i3}[c][c][1]{$y_{15}$}
\psfrag{i4}[c][c][1]{$y_{17}$}
\psfrag{i5}[c][c][1]{$y_{18}$}
\psfrag{0}[c][c][1]{$0$}
\psfrag{a}[c][c][1]{$x_{14|r}$}
\psfrag{r1}[c][c][1]{$r$}
\psfrag{r2}[c][c][1]{$r$}
\psfrag{r3}[c][c][1]{$r$}
\psfrag{t1}[c][c][1]{$t$}
\psfrag{t2}[c][c][1]{$t$}
\psfrag{t3}[c][c][1]{$t$}
\psfrag{t4}[c][c][1]{$r$}
\psfrag{t5}[c][c][1]{$t$}
\psfrag{x2:}[c][c][1]{$\mathbf{x}_2$\textbf{:}}
\psfrag{u:}[c][c][1]{$\mathbf{u}$\textbf{:}}
\psfrag{x1:}[c][c][1]{$\mathbf{x}_1$\textbf{:}}
\psfrag{y1:}[c][c][1]{$\mathbf{y}_1$\textbf{:}}
\psfrag{x1r:}[c][c][1]{$\mathbf{x}_{1|r}$\textbf{:}}
\psfrag{y1r:}[c][c][1]{$\mathbf{y}_{1|r}$\textbf{:}}
\psfrag{y2:}[c][c][1]{$\mathbf{y}_{2}$\textbf{:}}
\psfrag{a1}[c][c][1]{$y_{21}$}
\psfrag{a2}[c][c][1]{$y_{22}$}
\psfrag{a3}[c][c][1]{$y_{23}$}
\psfrag{a4}[c][c][1]{$y_{24}$}
\psfrag{a5}[c][c][1]{$y_{25}$}
\psfrag{a6}[c][c][1]{$y_{26}$}
\psfrag{a7}[c][c][1]{$y_{27}$}
\psfrag{a8}[c][c][1]{$y_{28}$}
}}
\caption{Example of  generated   switching vector along with input/output codewords at   source, relay, and destination. }  \label{fig_x2_tocna}
\end{figure}

In Fig.~\ref{fig_x2_tocna}, we show
an example for  vectors $\mathbf{x}_{1|r}$, $\mathbf{x}_{1}$,  $\mathbf{y}_{1}$,  $\mathbf{y}_{1|r}$, $\mathbf{x}_2$, and $\mathbf{y}_2$,     for $k=8$ and $P_U^*=1/2$, where   $\mathbf{x}_1$   contains   all $k$ input symbols  at the source  including the silences. From this example, it can be seen that $\mathbf{x}_1$   contains zeros due to silences for channel uses for  which the corresponding symbol in $\mathbf{x}_2$ is non-zero.  By comparing    $\mathbf{x}_1$  and $\mathbf{x}_2$ it can be seen that the HD constraint is satisfied for each symbol duration.

\begin{figure}
\centering
 \resizebox{1\linewidth}{!}{
\pstool[width=1\linewidth]{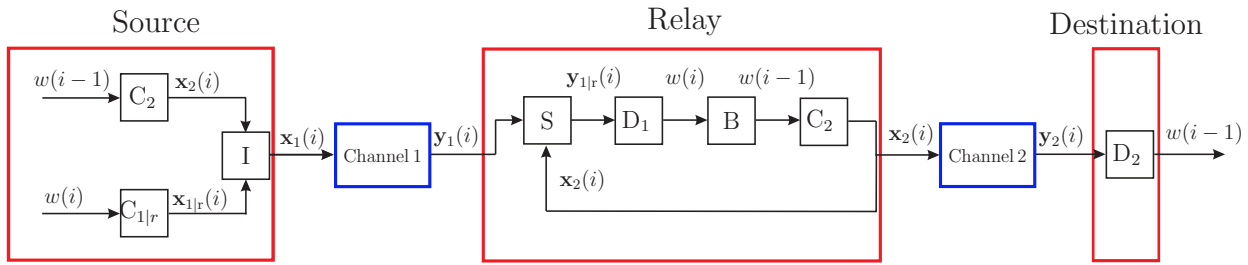}{
\psfrag{w1}[c][c][0.7]{$w(i-1)$}
\psfrag{w2}[c][c][0.7]{$w(i)$}
\psfrag{u1}[c][c][0.7]{$\mathbf{u}(i)$}
\psfrag{xs}[c][c][0.7]{$\mathbf{x}_{1|\mathrm{r}}(i)$}
\psfrag{x1}[c][c][0.7]{$\mathbf{x}_1(i)$}
\psfrag{y1}[c][c][0.7]{$\mathbf{y}_1(i)$}
\psfrag{yr}[c][c][0.7]{$\mathbf{y}_{1|\mathrm{r}}(i)$}
\psfrag{x2}[c][c][0.7]{$\mathbf{x}_2(i)$}
\psfrag{y2}[c][c][0.7]{$\mathbf{y}_2(i)$}
\psfrag{1}[c][c][0.8]{$\mathrm{C}_{1|r}$}
\psfrag{2}[c][c][0.8]{$\mathrm{C}_2$}
\psfrag{U}[c][c][0.8]{$\mathrm{C}_2$}
\psfrag{M}[c][c][0.8]{$\mathrm{I}$}
\psfrag{E}[c][c][0.8]{$\mathrm{S}$}
\psfrag{3}[c][c][0.8]{$\mathrm{D}_1$}
\psfrag{4}[c][c][0.8]{$\mathrm{D}_2$}
\psfrag{B}[c][c][0.8]{$\mathrm{B}$}
\psfrag{C1}[c][c][0.6]{$\mathrm{Channel}\,1$}
\psfrag{C2}[c][c][0.6]{$\mathrm{Channel}\,2$}
\psfrag{Source}[c][c][1]{$\mathrm{Source}$}
\psfrag{Relay}[c][c][1]{$\mathrm{Relay}$}
\psfrag{Des}[c][c][1]{$\mathrm{Destination}$}
}}
\caption{Block diagram of the proposed channel coding protocol for time slot $i$. The following notations are used in the  block diagram: $\mathrm{C}_{1|r}$  and $\mathrm{C}_2$ are encoders, $\mathrm{D}_1$ and $\mathrm{D}_2$ are decoders,  $\mathrm{I}$ is an inserter,  $\mathrm{S}$ is a selector,  $\mathrm{B}$ is a buffer, and $w(i)$ denotes the message   transmitted by the source in block $i$.}
\label{Fig:Channel}
\end{figure}

The block diagram of the proposed coding scheme  is shown in Fig.~\ref{Fig:Channel}. In particular,
in Fig~\ref{Fig:Channel}, we show schematically the encoding,  transmission, and decoding at  source, relay, and destination. The flow of encoding/decoding in Fig.~\ref{Fig:Channel}    is as follows. Messages $w(i-1)$ and $w(i)$ are encoded into $\mathbf{x}_2(i)$ and $\mathbf{x}_{1|r}(i)$, respectively, at the source using the encoders  $\mathrm{C}_2$ and $\mathrm{C}_{1|r}$,  respectively. Then, an inserter $\mathrm{I}$  is used to create the vector $\mathbf{x}_{1}(i)$ by inserting the symbols of $\mathbf{x}_{1|r}(i)$ into the positions of $\mathbf{x}_{1}(i)$ for  which the corresponding elements of $\mathbf{x}_2(i)$ are zeros and setting all other symbols in $\mathbf{x}_{1}(i)$ to zero.
 The source then transmits $\mathbf{x}_{1}(i)$.  On the other hand, the relay,  encodes  $w(i-1)$ into  $\mathbf{x}_{2}(i)$ using encoder  $\mathrm{C}_2$. Then, the relay  transmits $\mathbf{x}_{2}(i)$ while receiving $\mathbf{y}_{1}(i)$. Next, using $\mathbf{x}_2(i)$, the relay constructs $\mathbf{y}_{1|r}(i)$ from $\mathbf{y}_{1}(i)$ by selecting only those symbols for which the corresponding symbol in  $\mathbf{x}_2(i)$ is zero. The relay then decodes $\mathbf{y}_{1|r}(i)$, using decoder $\mathrm{D}_1$, into $w(i)$ and stores the decoded bits in its buffer $\mathrm{B}$. The destination receives $\mathbf{y}_{2}(i)$, and decodes it using decoder $\mathrm{D}_2$, into $w(i-1)$.

\section{Capacity Examples}\label{sec_c-4}
In the following, we evaluate the capacity of the considered relay channel when the   source-relay and relay-destination links are both  BSCs and   AWGN channels, respectively.

\subsection{Binary Symmetric Channels}
Assume that the source-relay and relay-destination links are both  BSCs, where $\mathcal{X}_1=\mathcal{X}_2=\mathcal{Y}_1=\mathcal{Y}_2=\{0,1\}$, with probability of error $P_{\varepsilon 1}$ and $P_{\varepsilon 2}$, respectively.  Now, in order to obtain the capacity for this relay channel, according to Theorem~\ref{SItheo_1}, we first have to find $\max\limits_{p(x_{1}|x_2=0)} I\big(X_{1}; Y_{1}| X_2=0\big)$  and $\max\limits_{p_V(x_{2})} I(X_2;Y_{2})$.
 For the BSC, the expression  for $\max\limits_{p(x_{1}|x_2=0)} I\big(X_{1}; Y_{1}| X_2=0\big)$ is well known and given by \cite{cover}
 \begin{eqnarray}
     \max_{p(x_{1}|x_2=0)} I\big(X_{1}; Y_{1}| X_2=0\big)=1-H(P_{\varepsilon1}),\label{eq_kk1} 
\end{eqnarray}
where $H(P_{\varepsilon1})$ is the binary entropy function, which for probability $P$ is  defined as
\begin{eqnarray}\label{eq_nn11}
    H(P)=-P\log_2(P) -(1-P)\log_2(1-P).
\end{eqnarray}
The distribution that maximizes $I\big(X_{1}; Y_{1}| X_2=0\big)$ is also well known and given by \cite{cover}
\begin{eqnarray}\label{eq_qq1}
   p(x_1=0|x_2=0)=p(x_1=1|x_2=0)=\frac{1}{2} .
\end{eqnarray}
On the other hand, for the BSC,  the only symbol in the set $\mathcal{X}_{2T}$ is  symbol $1$, which RV $V$   takes with probability one. In other words, $p_V(x_2)$ is a degenerate distribution, given by $p_V(x_2)=\delta(x_2-1)$. Hence, 
\begin{align}\label{eq_n-skdks}
   \max\limits_{p_V(x_{2})} I(X_2;Y_{2}) &=   I(X_2;Y_{2}) \Big|_{p_V(x_2)=\delta(x_2-1)} \\
&=      H(Y_{2})\Big|_{p_V(x_2)=\delta(x_2-1)}   -H(Y_2|X_2)\Big|_{p_V(x_2)=\delta(x_2-1)}.
\end{align}
For the BSC, the expression  for   $H(Y_{2}|X_{2})$ is independent of $X_2$,  and  is given by  \cite{cover}
 \begin{eqnarray}
H(Y_{2}|X_{2})&=&H(P_{\varepsilon 2})\label{eq_kk2}.
\end{eqnarray} 
On the other hand, in order to find $H(Y_{2})\Big|_{p_V(x_2)=\delta(x_2-1)}$ from (\ref{eq_H(y2)}),  we need  the distributions of $p(y_{2}|x_2=0)$ and $p(y_{2}|x_2=1)$. For the BSC with probability of error $P_{\varepsilon 2}$, these distributions are obtained    as
\begin{eqnarray}\label{eq_BSC_py2_r}
 p(y_{2}|x_2=0)=\left\{
\begin{array}{ll}
1-P_{\varepsilon 2}&\textrm{if } y_2=0\\
P_{\varepsilon 2}&\textrm{if } y_2=1,
\end{array}
\right.
\end{eqnarray}
and
\begin{eqnarray}\label{eq_BSC_py2_t}
 p(y_{2}|x_2=1)=\left\{
\begin{array}{ll}
P_{\varepsilon 2}   &\textrm{if } y_2=0\\
  1-P_{\varepsilon 2}  &\textrm{if } y_2=1.\\
\end{array}
\right.
\end{eqnarray}
Inserting (\ref{eq_BSC_py2_r}),  (\ref{eq_BSC_py2_t}), and $p_V(x_2)=\delta(x_2-1) $ into (\ref{eq_H(y2)}), we obtain $H(Y_{2})\big|_{p_V(x_2)=\delta(x_2-1) }$ as
\begin{eqnarray}\label{eq_BSC_max_H(Y2)}
   H(Y_{2})\big|_{p_V(x_2)=\delta(x_2-1) } = 
-A\log_2(A)-(1-A)\log_2(1-A),  
\end{eqnarray}
where  
\begin{equation}\label{eq_A}
   A=P_{\varepsilon 2}(1-2 P_U)+P_U .
\end{equation}
Inserting (\ref{eq_kk2}) and (\ref{eq_BSC_max_H(Y2)}) into (\ref{eq_n-skdks}), we obtain $\max\limits_{p_V(x_2)} I(X_2;Y_{2}) $ as
\begin{align}\label{eq_BSC_max_I(X2;Y2)}
    \max_{ p_V(x_{2})} I(X_2;Y_2)= 
-A\log_2(A)-(1-A)\log_2(1-A) - H(P_{\varepsilon 2})  .
\end{align}

We now have the two necessary components required for obtaining  $P_U^*$ from (\ref{SIeq_capacity_2}), and thereby obtaining the capacity. This is summarized  in the following corollary.
\begin{corollary}\label{cor_1}
The capacity of the considered  relay channel with   BSCs links is given  by
 \begin{align}\label{eq_dnfj}
C=\max_{P_U} \min\big\{(1-H(P_{\varepsilon1}))(1-P_U),-A\log_2(A)-(1-A)\log_2(1-A) - H(P_{\varepsilon 2}) \big\}
\end{align}
and is achieved with
 \begin{align} 
 &  p_V(x_{2})=\delta(x_2-1) \label{eq_ww2}\\
&  p(x_1=0|x_2=0)= p(x_1=1|x_2=0)  = 1/2\label{eq_ww3}.
\end{align}
There are two cases for the   optimal $P_U^*$  which maximizes (\ref{eq_dnfj}).
 If $P_U$ found from\footnote{ Solving (\ref{eq_dnfj1}) with respect to $P_U$ leads to a nonlinear equation, which can be easily solved using e.g. Newton's method \cite{deuflhard2011newton}.}
\begin{align} \label{eq_dnfj1}
(1-H(P_{\varepsilon1}))(1-P_U) = -A\log_2(A)-(1-A)\log_2(1-A) - H(P_{\varepsilon 2})
\end{align}
is smaller than $1/2$, then the optimal   $P_U^*$  which maximizes (\ref{eq_dnfj}) is found as the solution to (\ref{eq_dnfj1}), and the capacity simplifies to
 \begin{align}\label{eq_dnfj2}
C= (1-H(P_{\varepsilon1}))(1-P_U^*) = -A^*\log_2(A^*)-(1-A^*)\log_2(1-A^*) - H(P_{\varepsilon 2})  ,
\end{align} 
where $A^*=A|_{P_U=P_U^*}$.
Otherwise, if $P_U$ found from (\ref{eq_dnfj1}) is $P_U\geq 1/2$, then the optimal $P_U^*$ which maximizes (\ref{eq_dnfj}) is $P_U^*=1/2$, and the capacity simplifies to
\begin{eqnarray}\label{eq_cap_BSC_1}
    C= 1-H(P_{\varepsilon 2}).
\end{eqnarray} 
\end{corollary}
\begin{IEEEproof}
The capacity in (\ref{eq_dnfj}) is obtained by inserting (\ref{eq_kk1}) and (\ref{eq_BSC_max_I(X2;Y2)}) into (\ref{SIeq_capacity_2}). On the other hand, for the BSC, the solution of   (\ref{eq_PU''}) is $P_U''=1/2$, whereas (\ref{eq_PU'}) simplifies to (\ref{eq_dnfj1}). Hence, using Theorem~\ref{SItheo_1}, we obtain that if $P_U'\leq P_U''=1/2$, then $P_U^*=P_U'$, where $P_U'$ is found from (\ref{eq_dnfj1}), in which case the capacity is given by (\ref{SIeq_capacity_2b}), which simplifies to (\ref{eq_dnfj2}) for the BSC. On the other hand,  if $P_U'> P_U''=1/2$, then  $P_U^*=P_U''=1/2$,    in which case the capacity is given by (\ref{SIeq_capacity_2a}), which simplifies to (\ref{eq_cap_BSC_1}) for the BSC.
\end{IEEEproof}


\subsection{AWGN Channels}\label{Sec-awgn}

In this subsection, we assume that the source-relay and relay-destination links are AWGN channels, i.e., channels which are impaired by  independent, real-valued, zero-mean AWGN with variances $\sigma_1^2$ and $\sigma_2^2$, respectively. More precisely, the outputs at the relay and the destination are given by
\begin{eqnarray}\label{eq_y2_expression}
    Y_k=X_k+N_k,\quad k\in\{1,2\},
\end{eqnarray}
where $N_k$ is a zero-mean Gaussian  RV with variance $\sigma_k^2$, $k\in\{1,2\}$, with distribution $p_{N_k}(z)$,  $k\in\{1,2\}$, $-\infty\leq z\leq\infty$.
Moreover, assume that the  symbols transmitted by the source  and the relay must satisfy the following average power constraints\footnote{If the optimal distributions $p(x_1|x_2=0)$ and $p_{V}(x_{2})$ turn out to be continuous,  the sums in (\ref{eq_cond_awgn_power}) should be replaced by integrals. We note however that the generalization of  capacity expressions for a discrete to a continuous channel model may not be always straightforward. An exception to this is the AWGN channel which has been well studied in the literature \cite{cover2012elements}.} 
\begin{equation}\label{eq_cond_awgn_power}
   \sum_{x_1\in\mathcal{X}_1} x_1^2\; p(x_1|x_2=0)\leq P_1  \;\textrm{ and } \;\sum_{x_2\in\mathcal{X}_{2T}} x_2^2\; p_{V}(x_{2}) \leq P_2.
\end{equation}
Obtaining the capacity for this relay channel using Theorem~\ref{SItheo_1},   requires expressions for the functions $\max\limits_{p(x_{1}|x_2=0)} I\big(X_{1}; Y_{1}| X_2=0\big)$ and $\max\limits_{p_V(x_{2})} I(X_2;Y_{2})= \max\limits_{p_V(x_{2})} \big[H(Y_{2}) - H(Y_{2}|X_2) \big]$. For the AWGN channel, the expressions for the mutual information $\max\limits_{p(x_{1}|x_2=0)} I\big(X_{1}; Y_{1}| X_2=0\big)$  and the entropy $H(Y_{2}|X_{2})$ are   well known and given by
\begin{eqnarray}
  \max\limits_{p(x_{1}|x_2=0)} I\big(X_{1}; Y_{1}| X_2=0\big) &=& \frac{1}{2} \log_2\left(1+\frac{P_1}{\sigma_1^2}\right)\label{eq_qw1}\\
 H(Y_2|X_2)&=& \frac{1}{2} \log_2\left(2\pi e \sigma_2^2\right),\label{eq_qw2}
\end{eqnarray}
where, as is well known, for AWGN  $I(X_1;Y_1|X_2=0)$ is maximized when $p(x_{1}|x_2=0)$ is the zero mean Gaussian distribution with variance $P_1$. On the other hand, $H(Y_2|X_2)$ is just the differential entropy of    Gaussian RV $N_2$, which is independent   of $p(x_2)$, i.e., of $p_V(x_{2})$. Hence, 
\begin{eqnarray}\label{dskdjks}
      \max\limits_{p_V(x_{2})} I(X_2;Y_{2})= \max\limits_{p_V(x_{2})}  H(Y_{2}) - \frac{1}{2} \log_2\left(2\pi e \sigma_2^2\right) 
\end{eqnarray}
 holds and in order to find $\max\limits_{p_V(x_{2})} I(X_2;Y_{2})$ we only need to derive $\max\limits_{p_V(x_{2})}  H(Y_{2})$.
 Now, in order to find $\max\limits_{p_V(x_{2})}  H(Y_{2})$, we first  obtain $H(Y_2)$ using (\ref{eq_H(y2)}) and then obtain  the distribution $p_V(x_{2})$ which maximizes $H(Y_2)$.
 Finding an expression for $H(Y_2)$ requires the distribution of $p(y_{2}|x_2)$. This distribution  is    found using (\ref{eq_y2_expression})   as
\begin{equation}\label{eq_pdf1a1}
    p(y_2|x_2)=p_{N_2}(y_2-x_2).
\end{equation}
 Inserting    (\ref{eq_pdf1a1})   into (\ref{eq_H(y2)}), we obtain $H(Y_2)$ as
\begin{align}\label{eq_sadfsdfs}
    H(Y_2) &= -\int_{-\infty}^\infty \bigg[ P_{U} \sum_{x_{2}\in\mathcal{X}_{2T}}  p_{N_2}(y_2-x_2) p_{V}(x_{2})  + p_{N_2}(y_2) (1-P_{U}) \bigg] \nonumber\\
&\qquad\times \log_2\bigg[ P_{U}  \sum_{x_{2}\in\mathcal{X}_{2T}}    p_{N_2}(y_2-x_2) p_{V}(x_{2}) + p_{N_2}(y_2) (1-P_{U}) \bigg]dy_2, 
\end{align}
where, since  $p(y_2|x_2)$ is now a continuos probability density function, the summation in (\ref{eq_H(y2)}) with respect to $y_2$ converges to an integral as 
\begin{eqnarray}\label{eq_pdf11a}
    \sum_{y_2}\rightarrow\int\limits_{-\infty}^\infty dy_2.
\end{eqnarray} 
We are now ready to maximize  $H(Y_2)$ in (\ref{eq_sadfsdfs}) with respect to $p_V(x_2)$. Unfortunately, obtaining the optimal  $p_V(x_2)$ which maximizes $H(Y_2)$ in closed form is difficult, if not impossible. However, as will be shown in the following lemma, we still can characterize the optimal  $p_V(x_2)$, which is helpful for numerical calculation of $p_V(x_2)$.
\begin{lemma}\label{lem_finite_distribution}
For the considered relay channel where the relay-destination link is an  AWGN channel and  where the input symbols of the relay must satisfy the average power constraint given in (\ref{eq_cond_awgn_power}),   the distribution $p_V(x_2)$ which maximizes $H(Y_2)$ in (\ref{eq_sadfsdfs}) for a fixed $P_U<1$  is discrete, symmetric around zero, and with infinite number of mass points, where the probability mass points in any bounded interval is finite, i.e., $p_V(x_2)$  has the following form
\begin{eqnarray}\label{eq_distrib_p(x_2|t)}
    p_V(x_2)=\sum_{k=1}^\infty p_k \delta(x_2-x_{2k}),
\end{eqnarray}
where $p_k$ is the probability that symbol $x_2$ will take the value $x_{2k}$, for $k=1,...,\infty$. Furthermore, $p_k$ and $x_{2k}$ given in (\ref{eq_distrib_p(x_2|t)}), must satisfy 
\begin{eqnarray}\label{eq_cond_2}
    \sum_{k=1}^\infty p_k =1\quad \textrm{and}\quad \sum_{k=1}^\infty p_k x_{2k}^2=P_2.
\end{eqnarray}
In the limiting case  when $P_U\to 1$,  distribution $p_V(x_2)$  converges to the zero-mean Gaussian distribution with variance $P_2$.
\end{lemma}
\begin{IEEEproof}
Please see Appendix~\ref{app_finite_distribution}.
\end{IEEEproof}
\begin{remark}
Unfortunately, there is no closed-form expression  for  distribution $p_V(x_2)$ given in the form of (\ref{eq_distrib_p(x_2|t)}), and therefore,   a  brute-force search  has to be   used in order to find $x_{2k}$ and $p_k$, $\forall k$. 
\end{remark}

Now, inserting   (\ref{eq_distrib_p(x_2|t)}) into (\ref{eq_sadfsdfs})
 we obtain 
$\max\limits_{p_V(x_2)} H(Y_2)$   as
\begin{eqnarray}\label{eq_H2_awgn}
  \max\limits_{p_V(x_2)} H(Y_2)=- \int\limits_{-\infty}^\infty &&\hspace{-5mm} \left( P_U\; \sum_{k=1}^\infty p_k^* p_{N_2}(y_2-x_{2k}^*) + (1-P_U) p_{N_2}(y_2)  \right) \nonumber\\
&&\times \log_2\left( P_U\; \sum_{k=1}^\infty p_k^* p_{N_2}(y_2-x_{2k}^*) + (1-P_U) p_{N_2}(y_2)  \right)   d y_2, 
\end{eqnarray}
where $p_V^*(x_2)=\sum_{k=1}^\infty p_k^* \delta(x_2-x_{2k}^*)$ is the distribution that maximizes $H(Y_2)$ in  (\ref{eq_sadfsdfs}). Inserting (\ref{eq_H2_awgn}) into (\ref{dskdjks}), we obtain $\max\limits_{p_V(x_2)} I(X_2;Y_2)$.
Using (\ref{eq_qw1})  and $\max\limits_{p_V(x_2)} I(X_2;Y_2)$ in Theorem~\ref{SItheo_1}, we obtain the capacity of the considered relay channel with AWGN links. This is conveyed in the following corollary.
\begin{corollary}\label{cor_cap_gauss}
The capacity of the considered relay channel where the source-relay and relay-destination links are both  AWGN channels with noise variances $\sigma_1^2$ and $\sigma_2^2$, respectively, and where the average power constraints of the inputs of source and relay are given by (\ref{eq_cond_awgn_power}), is given by
\begin{eqnarray}\label{eq_cap_gauss}
    C &=& \frac{1}{2} \log_2\left(1+\frac{P_1}{\sigma_1^2}\right)(1-P_U^*)\nonumber\\
&\stackrel{(a)}{=}& - \int\limits_{-\infty}^\infty \left( P_U^*\; \sum_{k=1}^\infty p_k^* p_{N_2}(y_2-x_{2k}^*) + (1-P_U^*) p_{N_2}(y_2)  \right) \nonumber\\
&&\;\;\times \log_2\left( P_U^*\; \sum_{k=1}^\infty p_k^* p_{N_2}(y_2-x_{2k}^*) + (1-P_U^*) p_{N_2}(y_2)  \right)   d y_2 -\frac{1}{2}\log_2(2\pi e \sigma_2^2),\ 
\end{eqnarray}
where  
the optimal $P_U^*$ is found such that  equality $(a)$ in (\ref{eq_cap_gauss}) holds. The capacity in (\ref{eq_cap_gauss}) is achieved when $p(x_1|x_2=0)$ is the zero-mean Gaussian distribution with variance $P_1$ and $p_V^*(x_2)=\sum_{k=1}^\infty p_k^* \delta(x_2-x_{2k}^*)$ is a discrete distribution  which satisfies (\ref{eq_cond_2}) and maximizes $H(Y_2)$ given in (\ref{eq_H2_awgn}). 
\end{corollary}
\begin{IEEEproof}
The capacity in (\ref{eq_cap_gauss}) is obtained by inserting (\ref{eq_H2_awgn}) into (\ref{dskdjks}),  then inserting (\ref{dskdjks}) and (\ref{eq_qw1}) into (\ref{SIeq_capacity_2}), and finally maximizing with respect to $P_U$. For the maximization of the corresponding capacity  with respect to $P_U$, we note that $P_U'<P_U''=1$ always holds. Hence,  the capacity is given by (\ref{SIeq_capacity_2b}), which for the Gaussian case simplifies to (\ref{eq_cap_gauss}). To see that  $P_U''=1$, note the relay-destination channel is an AWGN channel for which the mutual information is maximized  when $p(x_2)$ is a Gaussian distribution. From (\ref{eq_p(x_2)}), we see that $p(x_2)$   becomes a Gaussian distribution if and only if $P_U=1$ and $p_V(x_2)$ also assumes a Gaussian distribution.
\end{IEEEproof}

\section{Numerical Examples}\label{sec_num}

In this section, we  numerically evaluate the capacities of the considered HD relay channel when the source-relay and relay-destination links are both BSCs  and AWGN channels, respectively. As a performance benchmark,   we   use the maximal achievable rate of conventional relaying \cite{cheap_paper}. Thereby, the source transmits to the relay one codeword with rate $\max\limits_{p(x_{1}|x_2=0)} I\big(X_{1}; Y_{1}| X_2=0\big)$ in $1-P_U$ fraction of the time, where $0<P_U<1$, and in the remaining  fraction of  time, $P_U$,  the relay retransmits the received information to the destination with rate $\max\limits_{p(x_2)}I(X_2;Y_2)$, see  \cite{1435648} and \cite{cheap_paper}. The optimal $P_U$, is found such that the following holds
\begin{align}\label{eq_conv1}
    R_{\rm conv}= \max\limits_{p(x_{1}|x_2=0)} I\big(X_{1}; Y_{1}| X_2=0\big)(1-P_U)  = \max_{p(x_2)}I(X_2;Y_2) P_U .
\end{align}
Employing the optimal $P_U$ obtained from (\ref{eq_conv1}), the maximal achievable rate of conventional relaying can be written as
\begin{eqnarray}\label{eq_cc}
    R_{\rm conv}
=\frac{\max\limits_{p(x_{1}|x_2=0)} I\big(X_{1}; Y_{1}| X_2=0\big) \times \max\limits_{p(x_2)}I(X_2;Y_2)   } {\max\limits_{p(x_{1}|x_2=0)} I\big(X_{1}; Y_{1}| X_2=0\big) + \max\limits_{p(x_2)}I(X_2;Y_2)  }.
\end{eqnarray}

\subsection{BSC Links}
\begin{figure} 
\centering
\includegraphics[width=6in]{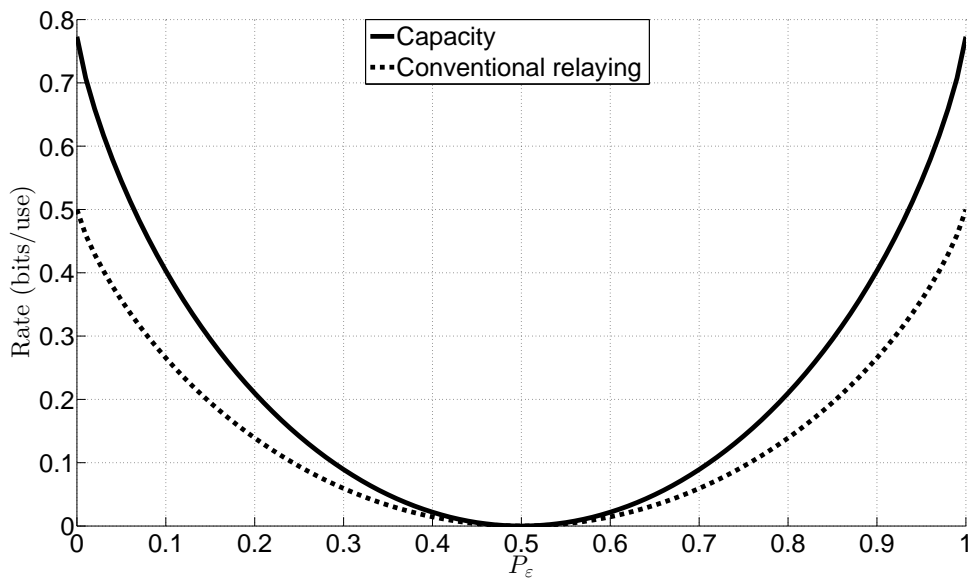}
\caption{Comparison of rates for the BSC as a function of the error probability  $P_{\varepsilon 1}=P_{\varepsilon 2}=P_{\varepsilon}$.}\label{fig_n1}
\end{figure}

For simplicity, we assume symmetric links with $P_{\varepsilon 1}=P_{\varepsilon 2}=P_{\varepsilon}$. As a result, $P_U^*<1/2$ in Corollary~\ref{cor_1}  and the capacity is given by (\ref{eq_dnfj2}). This capacity is plotted in Fig.~\ref{fig_n1}, where  $P_U^*$ is found from (\ref{eq_dnfj1}) using a  mathematical software package, e.g. Mathematica. As a benchmark, in Fig.~\ref{fig_n1}, we also show  the maximal achievable rate using conventional relaying, obtained  by inserting
\begin{eqnarray}
   \max\limits_{p(x_{1}|x_2=0)} I\big(X_{1}; Y_{1}| X_2=0\big)=\max\limits_{p(x_2)}I(X_2;Y_2)= 1- H(P_\varepsilon)
\end{eqnarray}
into  (\ref{eq_cc}), where $H(P_\varepsilon)$ is given in (\ref{eq_nn11}) with $P=P_\varepsilon$.
   Thereby, the following rate  is obtained 
\begin{equation}\label{eq_n2}
    R_{\rm conv}=\frac{1}{2}\big(1- H(P_\varepsilon)\big). 
\end{equation}
 As can be seen from Fig.~\ref{fig_n1}, when both links are error-free, i.e., $P_{\varepsilon}=0$,  conventional relaying achieves $0.5$ bits/channel use, whereas the capacity is  $0.77291$, which is   $54\%$ larger than the rate achieved with conventional relaying. This value for the capacity can be obtained by inserting $P_{\varepsilon 1}=P_{\varepsilon 2}=0$ in (\ref{eq_dnfj1}), and thereby obtain
\begin{equation}\label{eq_cap_BSC_a2}
    C= 
1-P_U^*  
\stackrel{(a)}{=} H(P_U^*). 
\end{equation} 
Solving $(a)$ in (\ref{eq_cap_BSC_a2}) with respect to $P_U^*$ and inserting the solution for $P_U^*$ back into   (\ref{eq_cap_BSC_a2}), yields $C=0.77291$. We note that this value was first reported in \cite[page 327]{kramer_book_1}. 

\subsection{AWGN Links}

\begin{figure}
\includegraphics[width=6in]{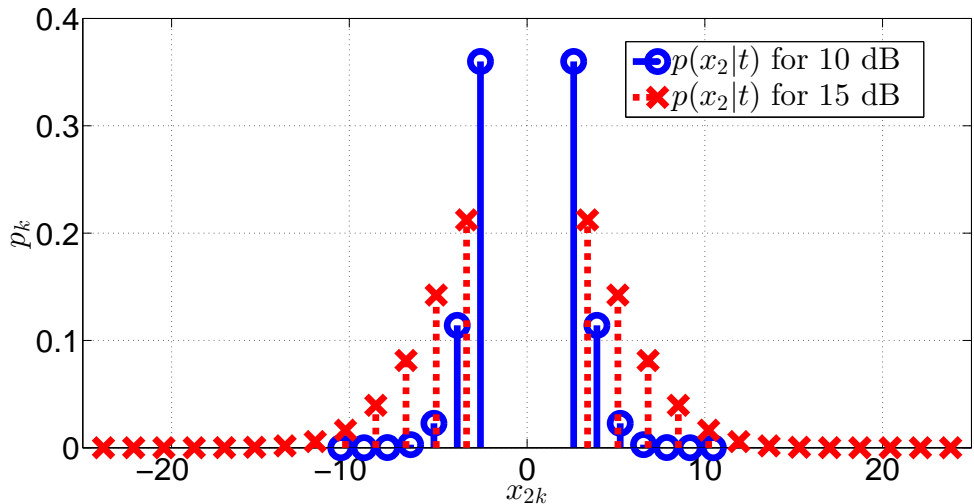}
\centering
\caption{Example of proposed input distributions at the relay $p_V(x_2)$.}\label{fig_dist_1}
\end{figure}

\begin{figure}
\includegraphics[width=6in]{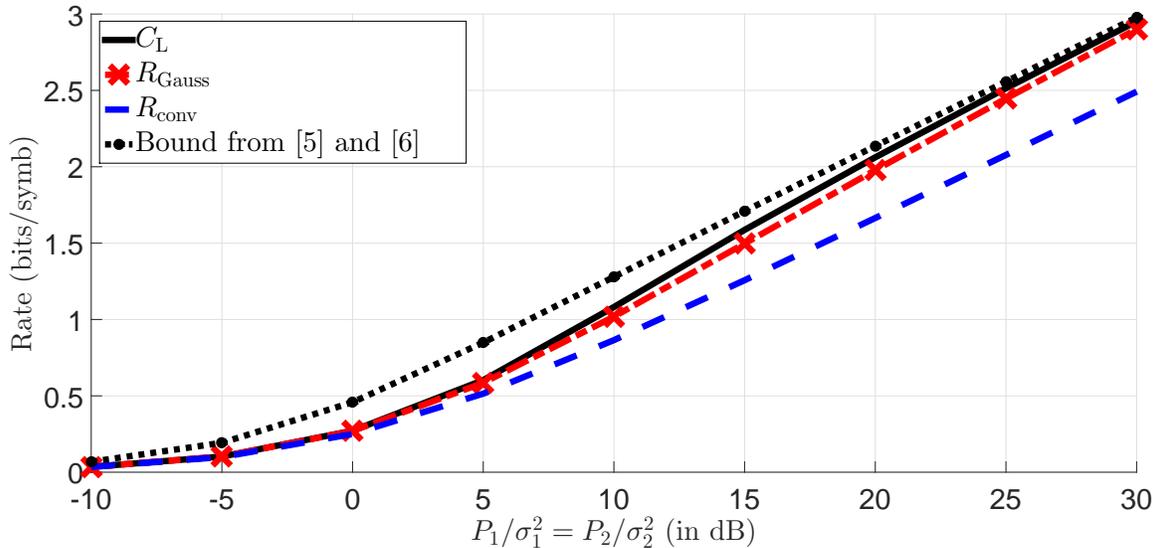}
\centering
\caption{Source-relay and relay destination links are AWGN channels with  $ P_1/\sigma_1^2= P_2/\sigma_2^2$.}\label{fig_n2}
\end{figure}

\begin{figure}
\includegraphics[width=6in]{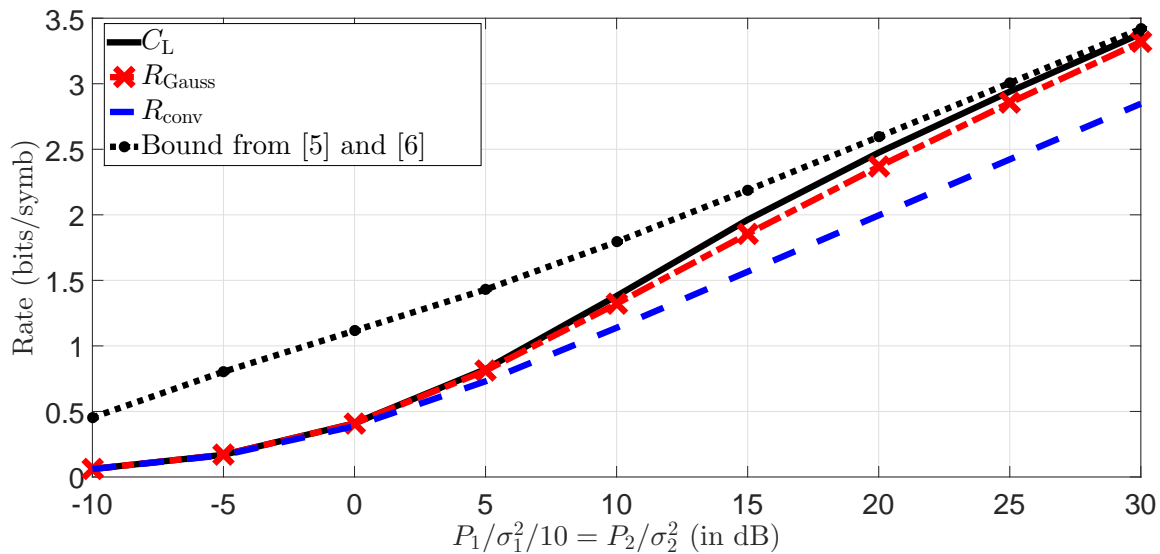}
\centering
\caption{Source-relay and relay destination links are AWGN channels with  $ P_1/\sigma_1^2/10= P_2/\sigma_2^2$.}\label{fig_n3}
\end{figure}

For the AWGN case,  the capacity is evaluated based on Corollary~\ref{cor_cap_gauss}. However, since for this case the optimal input distribution at the relay $p_V^*(x_2)$ is unknown, i.e., the values of $p_k^*$ and $x_{2k}^*$ in (\ref{eq_cap_gauss}) are unknown, we have performed a brute force search for the   values of $p_k^*$ and $x_{2k}^*$ which maximize (\ref{eq_cap_gauss}). Two examples of such distributions\footnote{ Note that these distributions resemble   a   discrete, Gaussian shaped distribution with a gap around zero.} are shown in Fig.~\ref{fig_dist_1} for two different values of the SNR $ P_1/\sigma_1^2= P_2/\sigma_2^2$. 
 Since we do not have a proof that the  distributions obtained via brute-force search are actually the exact optimal input distributions at the relay  that achieve  the capacity, the rates that we obtain,  denoted by $C_{\rm L}$,  are lower than or equal to the actual capacity.   These rates are shown in  Figs.~\ref{fig_n2} and \ref{fig_n3},  for symmetric links and non-symmetric links, respectively, where  we set $P_1/\sigma_1^2=P_2/\sigma_2^2$ and  $ P_1/\sigma_1^2/10= P_2/\sigma_2^2$, respectively. We note that for the results in Fig.~\ref{fig_n2},  for $P_1/\sigma_1^2=P_2/\sigma_2^2=10$ dB and  $P_1/\sigma_1^2=P_2/\sigma_2^2=15$ dB,  we have used the input distributions at the relay shown in Fig.~\ref{fig_dist_1}. In particular, for $P_1/\sigma_1^2=P_2/\sigma_2^2=10$ dB we have used the following values for $p_k^*$ and $x_{2k}^*$\\
$p_k^*=[ 
0.35996,\; 0.11408,\; 2.2832\times 10^{-2},\; 2.88578\times 10^{-3},\; 2.30336\times 10^{-4}, $

$\quad\;\; 1.16103\times 10^{-5},\; 
3.69578\times 10^{-7}],$
\\
$x_{2k}^*=[2.62031,\; 3.93046,\; 5.24061, 6.55077,\; 7.86092,\; 9.17107, \;10.4812],$\\
and for $P_1/\sigma_1^2=P_2/\sigma_2^2=15$ dB we have used\\
$p_k^*=[ 0.212303, 0.142311, 
8.12894\times 10^{-2}, 3.95678\times 10^{-2}, 1.64121\times 10^{-2}, 5.80092\times 10^{-3}$\\
$ 1.7472\times 10^{-3}, 4.48438\times 10^{-4}, 9.80788\times 10^{-5}, 1.82793\times 10^{-5}, 2.90308\times 10^{-6}, 3.92889\times 10^{-7}],$
\\
$x_{2k}^*=[3.40482,\; 
5.10724,\; 6.80965,\; 8.51206,\; 10.2145,\; 11.9169,\; 13.6193,\; 15.3217,$

$\qquad 
  17.0241, \;  18.7265,\; 20.4289,\; 22.1314].$

\noindent
The above values of $p_k^*$ and $x_{2k}^*$ are only given for  $x_{2k}^*>0$, since the  values of $p_k^*$ and $x_{2k}^*$ when $x_{2k}^*<0$ can be found from  symmetry, see Fig.~\ref{fig_dist_1}.

In Figs.~\ref{fig_n2} and \ref{fig_n3}, we also show the rate achieved when instead of an optimal discrete input distribution at the relay $p_V^*(x_2)$, cf. Lemma~\ref{lem_finite_distribution},   we use  a continuous, zero-mean Gaussian distribution with variance $P_2$. Thereby,  we obtain the following rate
\begin{eqnarray} 
    R_{\rm Gauss} &=& \frac{1}{2} \log_2\left(1+\frac{P_1}{\sigma_1^2}\right)(1-P_U)\nonumber\\
&\stackrel{(a)}{=}& - \int\limits_{-\infty}^\infty \big( P_U\; p_{G}(y_2) + (1-P_U) p_{N_2}(y_2)  \big) \nonumber\\
&&\qquad\times \log_2 \big( P_U\; p_{G}(y_2) + (1-P_U) p_{N_2}(y_2)  \big)   d y_2 -\frac{1}{2}\log_2(2\pi e \sigma_2^2), 
\end{eqnarray}
 where $P_U$ is found such that equality $(a)$ holds and $p_{G}(y_2)$ is a continuous, zero-mean Gaussian distribution with variance $P_2+\sigma_2^2$. From  Figs.~\ref{fig_n2} and \ref{fig_n3}, we can see that $R_{\rm Gauss}\leq C_{\rm L} $, which was expected from Lemma~\ref{lem_finite_distribution}. However, the loss in performance caused by the Gaussian inputs is moderate, which suggests that the performance gains obtained by the proposed protocol are mainly due to the exploitation  of the silent (zero) symbols for conveying information from the HD relay to the destination rather than the optimization of $p_V(x_2)$.

As benchmark, in Figs.~\ref{fig_n2} and \ref{fig_n3}, we have also shown the maximal achievable rate using conventional relaying, obtained  by inserting
\begin{equation}
\max\limits_{p(x_{1}|x_2=0)} I\big(X_{1}; Y_{1}| X_2=0, U=r\big)=\frac{1}{2}\log_2\left(1+\frac{P_1}{\sigma_1^2}\right) 
\end{equation}
and
\begin{equation}
 \max\limits_{p_V(x_2)}I(X_2;Y_2|U=t) 
 = \frac{1}{2}\log_2\left(1+\frac{P_2}{\sigma_2^2}\right) 
\end{equation}
into  (\ref{eq_cc}), which yields
\begin{eqnarray}
    R_{\rm conv}=\frac{1}{2} \frac{\log_2\left(1+\frac{P_1}{\sigma_1^2}\right)  \log_2\left(1+\frac{P_2}{\sigma_2^2}\right)  }{ \log_2\left(1+\frac{P_1}{\sigma_1^2}\right) + \log_2\left(1+\frac{P_2}{\sigma_2^2}\right) } .
\end{eqnarray}
Comparing the rates $C_{\rm L}$ and $R_{\rm conv}$ in Figs.~\ref{fig_n2} and \ref{fig_n3}, we see that for $10 \textrm{ dB}\leq P_2/\sigma_2^2\leq 30$ dB,  $C_{\rm L}$ achieves   $3$ to $6$ dB gain compared to $R_{\rm conv}$. Hence, large performance gains are achieved using the proposed capacity protocol even if  suboptimal input distributions at the relay are employed.

Finally, as additional benchmark in Figs.~\ref{fig_n2} and \ref{fig_n3}, we  show the  upper  bounds, achievable to withing 1 bit/symbol, reported in \cite{6763001}  and \cite{kramer_book_1}, given by
\begin{equation}
     C_{\rm Upper} = \max_{P_U}\min\Bigg\{\frac{1}{2} \log_2\left(1+\frac{P_1}{\sigma_1^2}\right)(1-P_U)\;,\;  \frac{1}{2} \log_2\left(1+\frac{P_2}{\sigma_2^2}\right)P_U +H(P_U)\Bigg\}. 
\end{equation}
 As can be seen from Figs.~\ref{fig_n2} and \ref{fig_n3}, this bound is   loose for low SNRs but becomes tight  for high SNRs.

\section{Conclusion}\label{sec-conc_c}

We have derived an easy-to-evaluate  expression for the   capacity of the two-hop HD relay channel by simplifying previously derived expression for the converse.   Moreover, we have proposed an explicit coding scheme which   achieves the capacity. In particular, we showed that the capacity is achieved when the relay sends additional information  to the destination by using the zero symbol implicitly generated by the relay's silence  during reception.  Furthermore, we have evaluated the  capacity for the cases when both links are BSCs and AWGN channels, respectively.  From the numerical examples, we have observed that the capacity of the two-hop HD relay channel is significantly higher than the rates achieved with conventional relaying protocols.

\appendix

\subsection{Proof That the Probability of Error at the Relay Goes to Zero When (\ref{eq_d_1aa}) Holds}\label{app_proof_of_prob_error_1}
In order to prove that the relay can decode the source's codeword in block $b$, $\mathbf{x}_{1|r}(b)$, where $1\leq b\leq N$,  from the received codeword  $\mathbf{y}_{1|r}(b)$  when (\ref{eq_d_1aa}) holds, i.e., that the probability of error at the relay goes to zero as $k\to\infty$, we will follow  the ``standard" method in \cite[Sec.~7.7]{cover2012elements} for analyzing the probability of error for rates smaller than the capacity.
To this end, note that the length  of   codeword  $\mathbf{x}_{1|r}(b)$ is $k(1-P_U^*)$. On the other hand, the  length  of   codeword  $\mathbf{y}_{1|r}(b)$ is  identical to the number of zeros\footnote{For $b=1$, note that the number of zeros in $\mathbf{x}_2(1)$ is $k$. Therefore, for $b=1$, we only take into an account the first $k(1-P_U^*)$ zeros. As a result, the length of $\mathbf{y}_{1|r}(b)$ is also $k(1-P_U^*)$.} in relay's transmit codeword $\mathbf{x}_2(b)$. Since the   zeros in  $\mathbf{x}_2(b)$ are generated independently using a coin flip, the number of zeros, i.e., the length of $\mathbf{y}_{1|r}(b)$  is $k(1-P_U^*)\pm \varepsilon(b)$, where $\varepsilon(b)$ is a non-negative integer. Due to the strong law of large numbers,   the following holds
\begin{align}
\lim_{k\to\infty} \frac{\varepsilon(b)}{k}&=0\label{eq_eq_app_a_1as},\\
\lim_{k\to\infty} \frac{k(1-P_U^*)\pm \varepsilon(b)}{k(1-P_U^*) }&=1\label{eq_eq_app_a_2as},
\end{align}
i.e., for large $k$, the  length  of relay's received  codeword  $\mathbf{y}_{1|r}(b)$ is approximately $k(1-P_U^*)$.

Now,  for block $b$,  we define a set $\mathcal{R}(b)$ which contains the symbol  indices $i$ in  block $b$ for which the symbols in $\mathbf{x}_2(b)$ are zeros, i.e., for which   $X_{2i}=0$. Note that before the start of the transmission in block $b$, the relay knows $\mathbf{x}_2(b)$, thereby it knows a priori for which symbol  indices   $i$ in block $b$, $X_{2i}=0$ holds. Furthermore, note that 
\begin{eqnarray}
    |\mathcal{R}(b)|=k(1-P_U^*)\pm \varepsilon(b) 
\end{eqnarray}
holds, where $|\cdot|$ denotes cardinality of a set.  Depending on the relation between $|\mathcal{R}(b)|$ and $k(1-P_U^*)$, the relay has to  distinguishes between two cases for decoding $\mathbf{x}_{1|r}(b)$    from   $\mathbf{y}_{1|r}(b)$. In the first case $ |\mathcal{R}(b)|\geq k(1-P_U^*)$ holds whereas in the second case  $ |\mathcal{R}(b)|<k(1-P_U^*)$ holds. We first explain the decoding procedure for the first case. 

When  $ |\mathcal{R}(b)|=k(1-P_U^*)+ \varepsilon(b)\geq k(1-P_U^*)$ holds, the source can transmit the entire codeword $\mathbf{x}_{1|r}(b)$, which is comprised of $k(1-P_U^*)$ symbols, since there are enough zeros in codeword $\mathbf{x}_{2}(b)$. On the other hand, since for this case the received codeword  $\mathbf{y}_{1|r}(b)$ is comprised of $k(1-P_U^*)+ \varepsilon(b)$ symbols, and since for the last $\varepsilon(b)$ symbols in  $\mathbf{y}_{1|r}(b)$ the source is silent, the relay keeps from  $\mathbf{y}_{1|r}(b)$ only the first $k(1-P_U^*)$ symbols and discards the remanning $\varepsilon(b)$ symbols. In this way, the relay keeps only the received symbols which are the result of the transmitted symbols in $\mathbf{x}_{1|r}(b)$, and discards the rest of the symbols in $\mathbf{y}_{1|r}(b)$ for which the source is silent. Thereby, from $\mathbf{y}_{1|r}(b)$, the relay generates a new received codeword which we denote  by $\mathbf{y}^*_{1|r}(b)$. Moreover, let $\mathcal{R}_1(b)$ be a set which contains the symbol indices of the symbols comprising codeword $\mathbf{y}^*_{1|r}(b)$. Now, note that the lengths  of $\mathbf{x}_{1|r}(b)$ and $\mathbf{y}^*_{1|r}(b)$, and the cardinality of set $\mathcal{R}_1(b)$ are $k(1-P_U^*)$, respectively. Having created $\mathbf{y}^*_{1|r}(b)$ and $\mathcal{R}_1(b)$, we now use a jointly typical decoder for decoding $\mathbf{x}_{1|r}(b)$ from $\mathbf{y}^*_{1|r}(b)$.  In particular,  we define a jointly typical set $A_\epsilon^{|\mathcal{R}_1(b)|}$  as 
\begin{subequations}
\begin{align}
&A_\epsilon^{|\mathcal{R}_1(b)|}=\Bigg\{  (\mathbf{x}_{1|r},\mathbf{y}^*_{1|r})\in \mathcal{X}_1^{|\mathcal{R}_1(b)|}\times \mathcal{Y}_1^{|\mathcal{R}_1(b)|}:\nonumber\\
  &   \left|-\frac{1}{|\mathcal{R}_1(b)|} \sum_{i\in \mathcal{R}_1(b)} \log_2 p(x_{1i}|x_{2i}=0) - H(X_1|X_2=0) \right| \leq \epsilon ,\label{eq_Ae_1}\\
  &   \left|-\frac{1}{|\mathcal{R}_1(b)|} \sum_{i\in \mathcal{R}_1(b)} \log_2 p(y_{1i}|x_{2i}=0) - H(Y_1|X_2=0) \right| \leq \epsilon, \label{eq_Ae_2}\\
  &  \left|-\frac{1}{|\mathcal{R}_1(b)|} \sum_{i\in \mathcal{R}_1(b)}  \log_2 p(x_{1i},y_{1i}|x_{2i}=0) - H(X_1,Y_1|X_2=0) \right| \leq \epsilon \Bigg\},\label{eq_Ae_3}
\end{align}
\end{subequations}
where $\epsilon$ is a small positive number.
The    transmitted codeword $\mathbf{x}_{1|r}(b)$ is  successfully decoded from received codeword $\mathbf{y}_{1|r}^*(b)$  if and only if  $(\mathbf{x}_{1|r}(b), \mathbf{y}_{1|r}^*(b))\in A_\epsilon^{|\mathcal{R}_1(b)|}$  and no other codeword  $\mathbf{\hat x}_{1|r}$ from  codebook $\mathcal{C}_{1|r}$ is jointly typical with $ \mathbf{y}_{1|r}^*(b)$. In order to compute the probability of error, we define the following events
\begin{eqnarray}
    E_0=\{(\mathbf{x}_{1|r}(b), \mathbf{y}_{1|r}^*(b))\notin A_\epsilon^{|\mathcal{R}_1(b)|}\}
\textrm{ and }
    E_j=\{(\mathbf{\hat x}_{1|r}^{(j)}, \mathbf{y}_{1|r}^*) \in A_\epsilon^{|\mathcal{R}_1(b)|}\},
\end{eqnarray}
where $\mathbf{\hat x}_{1|r}^{(j)}$ is the $j$-th codeword in $\mathcal{C}_{1|r}$ that is different from $\mathbf{x}_{1|r}(b)$. Note that in $\mathcal{C}_{1|r}$ there are $|\mathcal{C}_{1|r}|-1=2^{k R}-1$  codewords that are different from $\mathbf{x}_{1|r}(b)$, i.e., $j=1,...,2^{k R}-1$. Hence, an error occurs if any of  the events $E_0$, $E_1$, ..., $E_{2^{k R}-1}$ occurs. Since $\mathbf{x}_{1|r}(b)$ is uniformly selected from the codebook  $\mathcal{C}_{1|r}$,   the average probability of error  is given by
\begin{eqnarray} \label{eq_P_error1}
    {\rm Pr}(\epsilon)= {\rm Pr}(E_0\cup E_1\cup...\cup E_{2^{k R}-1})\leq {\rm Pr}(E_0)+\sum_{j=1}^{2^{k R}-1} {\rm Pr}(E_j).
\end{eqnarray}
Since $|\mathcal{R}_1(b)|\to\infty$ as $k\to\infty$, ${\rm Pr}(E_0)$  in (\ref{eq_P_error1})  is upper bounded as \cite[Eq. (7.74)]{cover2012elements}
\begin{eqnarray}\label{eq_PE_0}
   {\rm Pr}(E_0)\leq \epsilon .
\end{eqnarray}
On the other hand, since $|\mathcal{R}_1(b)|\to\infty$ as $k\to\infty$,  $ {\rm Pr}(E_j)$ is upper bounded as
 \begin{eqnarray}\label{eq_PE_j}
   {\rm Pr}(E_j)&=&{\rm Pr}\left(\left(\mathbf{\hat x}_{1|r}^{(j)},\mathbf{y}_{1|r}^*(b)\right)\in A_\epsilon^{|\mathcal{R}_1(b)|} \right)=\sum_{(\hat{\mathbf{x}}_{1|r}^{(j)}, \mathbf{y}_{1|r}^*(b))\in A_\epsilon^{|\mathcal{R}_1(b)|}} p(\mathbf{\hat x}_{1|r}^{(j)},\mathbf{y}_{1|r}^*(b)) \nonumber\\
&\stackrel{(a)}{=}& \sum_{(\hat{\mathbf{x}}_{1|r}^{(j)}, \mathbf{y}_{1|r}^*(b))\in A_\epsilon^{|\mathcal{R}_1(b)|}} p(\mathbf{\hat x}_{1|r}^{(j)}) p(\mathbf{y}_{1|r}^*(b))\nonumber\\
&\stackrel{(b)}{\leq}& \sum_{(\hat{\mathbf{x}}_{1|r}^{(j)}, \mathbf{y}_{1|r}^*(b))\in A_\epsilon^{|\mathcal{R}_1(b)|}} 2^{-|\mathcal{R}_1(b)| (H(X_1|X_2=0)-\epsilon)} 2^{-|\mathcal{R}_1(b)| (H(Y_1|X_2=0)-\epsilon)}\nonumber\\
&=& |A_\epsilon^{|\mathcal{R}_1(b)|}| 2^{-|\mathcal{R}_1(b)| (H(X_1|X_2=0)-\epsilon)} 2^{-|\mathcal{R}_1(b)| (H(Y_1|X_2=0)-\epsilon)}
\nonumber\\
&\stackrel{(c)}{\leq}& 2^{|\mathcal{R}_1(b)| (H(X_1,Y_1|X_2=0)+\epsilon)}
 2^{-|\mathcal{R}_1(b)| (H(X_1|X_2=0)-\epsilon)}  
2^{-|\mathcal{R}_1(b)| (H(Y_1|X_2=0)-\epsilon)}
\nonumber\\
&=& 2^{-|\mathcal{R}_1(b)|(H(X_1|X_2=0) + H(Y_1|X_2=0)- H(X_1,Y_1|X_2=0)-3\epsilon)}\nonumber\\
&=& 2^{-|\mathcal{R}_1(b)| (I(X_1;Y_1|X_2=0)-3\epsilon)},\qquad
\end{eqnarray}
where
$(a)$ follows since $\mathbf{\hat x}_{1|r}^{(j)}$ and $\mathbf{y}_{1|r}^*(b)$ are independent, $(b)$ follows since 
$$p(\mathbf{\hat x}_{1|r}^{(j)})\leq  2^{-|\mathcal{R}_1(b)| (H(X_1|X_2=0)-\epsilon)}\;\;\textrm{ and }\;\;
p(\mathbf{y}_{1|r}^*(b))\leq  2^{-|\mathcal{R}_1(b)| (H(Y_1|X_2=0)-\epsilon)},$$
 which follows from \cite[Eq. (3.6)]{cover2012elements}, respectively, and   $(c)$ follows since $$ |A_\epsilon^{|\mathcal{R}_1(b)|}|\leq  2^{|\mathcal{R}_1(b)| (H(X_1,Y_1|X_2=0)+\epsilon)},$$ which follows from  \cite[Theorem 7.6.1]{cover2012elements}.
Inserting (\ref{eq_PE_0}) and  (\ref{eq_PE_j}) into (\ref{eq_P_error1}), we obtain 
\begin{eqnarray} \label{eq_P_error1a}
    {\rm Pr}(\epsilon)&\leq& \epsilon+\sum_{j=1}^{2^{k R}-1}  2^{-|\mathcal{R}_1(b)| (I(X_1;Y_1|X_2=0)-3\epsilon)} \nonumber\\
& \leq &\epsilon+(2^{k R}-1)  2^{-|\mathcal{R}_1(b)|(I(X_1;Y_1|X_2=0)-3\epsilon)} \nonumber\\
&\leq& \epsilon+2^{k R}  2^{-|\mathcal{R}_1(b)| (I(X_1;Y_1|X_2=0)-3\epsilon)}
\nonumber\\
&=&
\epsilon+  
  2^{-k((|\mathcal{R}_1(b)|/k)  I(X_1;Y_1|X_2=0)-R- 3(1-P_U^*) \epsilon)}. 
\end{eqnarray}
Hence,  if 
\begin{align}\label{asdsdsdfsd}
R &<\frac{|\mathcal{R}_1(b)|}{k}  I(X_1;Y_1|X_2=0)- 3(1-P_U^*) \epsilon\nonumber\\
&=(1-P_U^*) I(X_1;Y_1|X_2=0) - 3(1-P_U^*)\epsilon,
\end{align}
  then $\lim\limits_{\epsilon\to 0}\lim\limits_{k\to\infty}{\rm Pr}(\epsilon)=0$. This concludes the proof for case when $ |\mathcal{R}(b)|\geq k(1-P_U^*)$ holds. We now turn to case two when $ |\mathcal{R}(b)|< k(1-P_U^*)$ holds.

When  $ |\mathcal{R}(b)|=k(1-P_U^*)- \varepsilon(b)< k(1-P_U^*)$ holds, then  the source cannot transmit all of its $k(1-P_U^*)$ symbols comprising codeword $\mathbf{x}_{1|r}(b)$ since there are not enough zeros in codeword $\mathbf{x}_{2}(b)$. Instead, the relay transmits  only $ k(1-P_U^*)-\varepsilon(b)$ symbols of codeword $\mathbf{x}_{1|r}(b)$, and  we denote the resulting transmitted codeword by  $\mathbf{x}_{1|r}^*(b)$. Note that the length of codewords $\mathbf{x}_{1|r}^*(b)$ and $\mathbf{y}_{1|r}(b)$, and the cardinality of $\mathcal{R}(b)$ are all identical and equal to $k(1-P_U^*)-\varepsilon(b)$.
 In addition, let the relay generate  a  codebook $\mathcal{C}_{1|r}^*(b)$   by keeping only the first $k(1-P_U^*)-\varepsilon(b)$ symbols  from each codeword in codebook $\mathcal{C}_{1|r}$ and discarding the remaining   $\varepsilon(b)$ symbols in the corresponding codewords. Let us denote the codewords in $\mathcal{C}_{1|r}^*(b)$ by $\mathbf{x}_{1|r}^*$. Note that there is a unique one to one mapping from the codewords in  $\mathcal{C}_{1|r}^*(b)$ to the codewords in $\mathcal{C}_{1|r}(b)$ since when $k\to\infty$, $k(1-P_U^*)-\varepsilon(b)\to\infty$ also holds, i.e., the lengths of the codewords in $\mathcal{C}_{1|r}^*(b)$ and $\mathcal{C}_{1|r} $ are  of the same order due to (\ref{eq_eq_app_a_2as}). Hence, if the relay can decode $\mathbf{x}_{1|r}^*(b)$ from $\mathbf{y}_{1|r}(b)$, then using this unique mapping between $\mathcal{C}_{1|r}^*(b)$ and $\mathcal{C}_{1|r}(b)$, the relay can decode $\mathbf{x}_{1|r}(b)$ and thereby   decode the message $w(b)$ sent from the source.

Now, for decoding $\mathbf{x}_{1|r}^*(b)$ from $\mathbf{y}_{1|r}(b)$, we again use jointly typical decoding. Thereby, we define a jointly typical set $B_\epsilon^{|\mathcal{R}|}$  as 
\begin{subequations}
\begin{align}
&B_\epsilon^{|\mathcal{R}|}=\Bigg\{  (\mathbf{x}_{1|r}^*,\mathbf{y}_{1|r})\in \mathcal{X}_1^{|\mathcal{R}|}\times \mathcal{Y}_1^{|\mathcal{R}|}:\nonumber\\
  &   \left|-\frac{1}{|\mathcal{R}|} \sum_{i\in \mathcal{R}} \log_2 p(x_{1i}|x_{2i}=0) - H(X_1|X_2=0) \right| \leq \epsilon ,\label{eq_Be_1}\\
  &   \left|-\frac{1}{|\mathcal{R}|} \sum_{i\in \mathcal{R}} \log_2 p(y_{1i}|x_{2i}=0) - H(Y_1|X_2=0) \right| \leq \epsilon, \label{eq_Be_2}\\
  &  \left|-\frac{1}{|\mathcal{R}|} \sum_{i\in \mathcal{R}}  \log_2 p(x_{1i},y_{1i}|x_{2i}=0) - H(X_1,Y_1|X_2=0) \right| \leq \epsilon \Bigg\}.\label{eq_Ae_3-1}
\end{align}
\end{subequations}

Again,  the    transmitted codeword $\mathbf{x}_{1|r}^*(b)$ is  successfully decoded from received codeword $\mathbf{y}_{1|r}(b)$  if and only if  $(\mathbf{x}_{1|r}^*(b), \mathbf{y}_{1|r}(b))\in B_\epsilon^{|\mathcal{R}|}$  and no other codeword  $\mathbf{\hat x}_{1|r}^*$ from  codebook $\mathcal{C}_{1|r}^*$ is jointly typical with $ \mathbf{y}_{1|r}(b)$. In order to compute the probability of error, we define the following events
\begin{eqnarray}
    E_0=\{(\mathbf{x}_{1|r}^*(b), \mathbf{y}_{1|r}(b))\notin B_\epsilon^{| \mathcal{R}|}\}
\textrm{ and }
    E_j=\{(\mathbf{\hat x}_{1|r}^{*(j)}, \mathbf{y}_{1|r}(b)) \in  B_\epsilon^{| \mathcal{R}|} \},
\end{eqnarray}
where $\mathbf{\hat x}_{1|r}^{*(j)}$ is the $j$-th codeword in $\mathcal{C}_{1|r}^*$ that is different from $\mathbf{x}_{1|r}^*(b)$.   Note that in $\mathcal{C}_{1|r}^*$ there are $|\mathcal{C}_{1|r}^*|-1=2^{k R}-1$  codewords that are different from $\mathbf{x}_{1|r}^*(b)$, i.e., $j=1,...,2^{k R}-1$. Hence, an error occurs if any of  the events $E_0$, $E_1$, ..., $E_{2^{k R}-1}$ occurs. Now,
using a similar procedure as for case when $ |\mathcal{R}(b)|\geq k(1-P_U^*)$, it can be proved that if 
\begin{align}\label{asdsdsdfsdaa}
R &<\frac{|\mathcal{R}(b)|}{k}  I(X_1;Y_1|X_2=0)- 3(1-P_U^*) \epsilon\nonumber\\
&=(1-P_U^*) I(X_1;Y_1|X_2=0) -\frac{\varepsilon(b)}{k}I(X_1;Y_1|X_2=0)  - 3(1-P_U^*)\epsilon,
\end{align}
 then $\lim\limits_{\epsilon\to 0}\lim\limits_{k\to\infty}{\rm Pr}(\epsilon)=0$.  In  (\ref{asdsdsdfsdaa}), note that 
\begin{eqnarray}
    \lim_{k\to\infty}\frac{\varepsilon(b)}{k} I(X_1;Y_1|X_2=0)=0 
\end{eqnarray}
holds due to (\ref{eq_eq_app_a_1as}). This concludes the proof for the case when $|\mathcal{R}(b)|<k(1-P_U^*)$.

\subsection{Proof of Lemma~\ref{lem_finite_distribution}}\label{app_finite_distribution}

Lemma~\ref{lem_finite_distribution} is proven using the results from \cite{6690191}, where the authors  investigate the optimal input distribution that achieves the capacity of an AWGN channel with an average power constraint $P_2$ and duty cycle $q$, where $0<q<1$. A duty cycle $q$ means that from $n\to\infty$ symbol transmissions, at least $q$ symbols have to be zero, see \cite{6690191}. In other words, for the channel $Y_2=X_2+N_2$, the authors of \cite{6690191} solve the following optimization problem
\begin{eqnarray}\label{eq:17ab}
\begin{array}{rl}
 {\underset{p(x_2)}{\rm{Maximize: }}}& I(X_2;Y_2)=h(Y_2)-\frac{1}{2} \log_2\left(2\pi e \sigma_2^2\right)\\
\vspace{1mm}
{\rm{Subject\;\; to}} \quad {\textrm C1:}& E\{X_2^2\}\leq P_2,\\
  {\textrm C2:}& {\rm Pr}\{X_2=0\}\geq q.
\end{array}
\end{eqnarray}
Now, since $\frac{1}{2} \log_2\left(2\pi e \sigma_2^2\right)$ is independent of  $p(x_2)$, ${\rm Pr}\{X_2=0\}\geq q$ is equivalent to ${\rm Pr}\{X_2\neq 0\}=P_U\leq 1-q$, and  $p(x_2)$ is given by $(\ref{eq_p(x_2)})$, the optimization problem in (\ref{eq:17ab}) can be written equivalently as
\begin{eqnarray}\label{eq:17ab_1}
\begin{array}{rl}
 {\underset{p_V(x_2),\;P_U}{\rm{Maximize: }}}& h(Y_2)\\
\vspace{1mm}
{\rm{Subject\;\; to}} \quad  {\textrm C1:}& E\{X_2^2\}\leq P_2,\\
  {\textrm C2:}& P_U\leq 1-q.
\end{array}
\end{eqnarray}
The authors in \cite{6690191} prove that solving   (\ref{eq:17ab_1}) for $q>0$ yields a discrete distribution for $p_V(x_2)$, symmetric around zero, and with infinite number of mass points, where the probability mass points in any bounded interval is finite, see Theorem~2 in \cite{6690191}.

On the other hand, the optimization problem that we need to solve in order to prove Lemma~\ref{lem_finite_distribution} is
\begin{eqnarray}\label{eq:17ab_2}
\begin{array}{rl}
 {\underset{p_V(x_2),\; P_U}{\rm{Maximize: }}}& h(Y_2)\\
\vspace{1mm}
{\rm{Subject\;\; to}} \quad  {\textrm C1:}& E\{V_2^2\}\leq P_2,\\
  {\textrm C2:}& P_U= 1- q.
\end{array}
\end{eqnarray}
Since  $p(x_2)$ is given by $(\ref{eq_p(x_2)})$,  $E\{V_2^2\}=E\{X_2^2\}$ holds. As a result, optimization problem  (\ref{eq:17ab_2}) can be written equivalently as
\begin{eqnarray}\label{eq:17ab_3}
\begin{array}{rl}
 {\underset{p_V(x_2),\; P_U}{\rm{Maximize: }}}& h(Y_2)\\
\vspace{1mm}
{\rm{Subject\;\; to}} \quad  {\textrm C1:}& E\{X_2^2\}\leq P_2,\\
  {\textrm C2:}& P_U= 1- q.
\end{array}
\end{eqnarray}
Now, since we can always increase $q$ in (\ref{eq:17ab_1}) such that constraint C2 in (\ref{eq:17ab_1}) holds with equality and since in that case again
 the optimal $p_V(x_2)$ of (\ref{eq:17ab_1})  is  discrete, symmetric around zero, and with infinite number of mass points, where the probability mass points in any bounded interval is finite, we obtain that
 the  optimal $p_V(x_2)$ of (\ref{eq:17ab_3}) also has to be discrete, symmetric around zero, and with infinite number of mass points, where the probability mass points in any bounded interval is finite. This concludes the proof of Lemma~\ref{lem_finite_distribution}.

\bibliography{litdab}
\bibliographystyle{IEEEtran}
\end{document}